\documentclass[fleqn,usenatbib]{mnras}

\usepackage{newtxtext,newtxmath}

\usepackage[T1]{fontenc}

\DeclareRobustCommand{\VAN}[3]{#2}
\let\VANthebibliography\thebibliography
\def\thebibliography{\DeclareRobustCommand{\VAN}[3]{##3}\VANthebibliography}

\usepackage{longtable}
\usepackage{color}
\newcommand{\project}[1]{\textsl{#1}}
 
\newcommand{\tc}{\project{The~Cannon}} 
\newcommand{\apogee}{\project{\textsc{apogee}}}
\newcommand{\lamost}{\project{\textsc{lamost}}}

\newcommand{\Gaia}{\project{Gaia}}

\newcommand{\galah}{\project{\textsc{galah}}}

\newcommand{\teff}{\mbox{$T_{\rm eff}$}}

\newcommand{\feh}{\mbox{$\rm [Fe/H]$}}
\newcommand{\mgfe}{\mbox{$\rm [Mg/Fe]$}}

\newcommand{\logg}{\mbox{$\log g$}}

\newcommand{\pampelmuse}{\project{PampelMUSE}}
\newcommand{\pmp}{\project{pymusepipe}}

\usepackage{multirow}
\usepackage{multicol}

\usepackage{graphicx}	% Including figure files
\usepackage{amsmath}

\title[ACACIAS I]{ACACIAS I: Element abundance labels for 192 stars in the dwarf galaxy NGC~6822}

\author[M. K. Ness et al.]{Melissa K. Ness,$^{1,2}$\thanks{E-mail: melissa.ness@anu.edu.au}
J. Trevor Mendel,$^{1,3}$
Sven Buder,$^{1,3}$
Adam Wheeler,$^{4}$
Alexander P. Ji,$^{5,6,7}$
Luka Mijnarends,$^{1,3}$ \newauthor
Kim Venn,$^{8}$
Else Starkenburg,$^{9}$
Ryan Leaman,$^{10}$
Kathryn Grasha,$^{1,3}$ and
Sarah Aquilina$^{1}$
\\
$^{1}$Research School of Astronomy \& Astrophysics, Australian National University, Canberra ACT 2611, Australia\\
$^{2}$Columbia University, Pupin Laboratories, New York City 11027\\
$^{3}$Center of Excellence for Astrophysics in Three Dimensions (ASTRO-3D), Australia\\
$^{4}$Department of Astronomy, Ohio State University, McPherson Laboratory, 140 West 18th Avenue, Columbus, Ohio, USA\\
$^{5}$Department of Astronomy \& Astrophysics, University of Chicago, 5640 S Ellis Avenue, Chicago, IL 60637\\
$^{6}$Kavli Institute for Cosmological Physics, University of Chicago, Chicago, IL 6063\\
$^{7}$Joint Institute for Nuclear Astrophysics—Center for the Evolution of the Elements (JINA-CEE), USA\\
$^{8}$Department of Physics \& Astronomy, University of Victoria, Victoria, BC, V8W 3P2, Canada\\
$^{9}$Kapteyn Astronomical Institute, University of Groningen, Postbus 800, NL9700 AV, Groningen, the Netherlands\\
$^{10}$Department of Astrophysics, University of Vienna, T\"{u}rkenschanzstra\ss{}e 17, 1180 Vienna, Austria
}

\date{Accepted YYYY Month DD. Received YYYY Month DD}

\pubyear{2024}

\begin{document}
\label{firstpage}
\pagerange{\pageref{firstpage}--\pageref{lastpage}}
\maketitle

\begin{abstract}
The element abundances of local group galaxies connect enrichment mechanisms to galactic properties and serve to contextualise the Milky Way's abundance distributions. Individual stellar spectra in nearby galaxies can be extracted from Integral Field Unit (IFU) data, and provide a means to take an abundance census of the local group. We introduce a program that leverages $R=1800$, $\mathrm{SNR}=15$, IFU resolved spectra from the Multi Unit Spectroscopic Explorer (MUSE). We deploy the data-driven modelling approach for labelling stellar spectra with stellar parameters and abundances, of \tc, on resolved stars in NGC~6822. We construct our model for \tc\ using $\approx$19,000 Milky Way 
\lamost\ spectra with \apogee\ labels. We report six \text{inferred} abundance labels (denoted $\ell_\mathrm{X}$), for 192  NGC~6822 disk stars, precise to $\approx$$0.15$ dex.  We validate our generated spectral models provide a good fit the data, including at individual atomic line features. We infer mean abundances of $\ell_\mathrm{[Fe/H]} = -0.90 \pm 0.03$, $\ell_\mathrm{[Mg/Fe]} = -0.01 \pm 0.01$, $\ell_\mathrm{[Mn/Fe]} = -0.22 \pm 0.02$, $\ell_\mathrm{[Al/Fe]} = -0.33 \pm 0.03$, $\ell_\mathrm{[C/Fe]} =-0.43 \pm 0.03$, $\ell_\mathrm{[N/Fe]} =0.18 \pm 0.03$. These abundance labels are similar to dwarf galaxies observed by \apogee, and the lower enhancements for NGC~6822 compared to the Milky Way are consistent with expectations. This approach supports a new era in extra-galactic archaeology of characterising the local group enrichment diversity using low-resolution, low-SNR IFU resolved spectra.
\end{abstract}


\section{Introduction} \label{sec:intro}

The growth of spectroscopic data of the Milky Way has launched an industry in Galactic archaeology. Measurements of stellar parameters,  velocities, element abundances, and ages of stars serve as the information to reconstruct Galactic history \citep[see][and references therein]{BH2016, Helmi2020, Deason2024}. \textit{Gaia} \citep{Gaia2018} has transformed the empirical characterisation of the Milky Way, on many scales. This includes in showcasing the diversity of the stellar halo's building blocks as well as the signatures of dynamical perturbations in the disk \citep[e.g.,][]{Naidu2020, Horta2023, Antoja2023, Hunt2022}. Ground-based spectroscopic surveys including Milky Way Mapper \citep{Kollmeier2017}, \apogee\ \citep{Majewski2017}, \galah\ \citep{deSilva2015} \textit{Gaia}-ESO \citep{Gilmore2012}, RAVE \citep{Steinmetz2006} and \lamost\ \citep{Zhang2023} are enabling a nucleosynthetic inventory of the Galaxy and a mapping of the relationship between ages, orbits and abundances \citep[e.g.,][]{Blancato2019, Sharma2022, Buder2023, Eilers2022, Molero2023, Hawkins2023, Lu2022, Patil2023, Ratcliffe2021, Queiroz2023, Xiang2022}.

Simulations are employed to interpret these data \citep[e.g.][]{ Loebman2016, Agertz, Roskar2008, Ratcliffe2024, Lu2022, Carrillo2023, Bell2022, Debattista2024, McCl2024}. However, there are many outstanding questions pertaining to the distribution of element abundances in the Milky Way.  The origins of the populations of the high and low-$\alpha$ element disk sequences at [Fe/H] $>$ $-1.0$ is a particularly significant feature that is debated. The two $\alpha$-sequences across [Fe/H] show different mean radial, age and dynamical properties and a number of contrasting explanations have been proposed. The [Fe/H]-$\alpha$ bi-modality may manifest from a single population with delayed chemical evolution, from radial migration, or derive from metal-rich gas inflow \citep{Sharma2021, Chen2023, IC2024, Chandra2023}. This characteristic may be present in other Galaxies \citep{Scott2021}. Irrespective of a star's membership in the low- and high-$\alpha$ sequence, the underlying dimensionality of the abundance measurements from current large stellar surveys appears to be low \citep[e.g.,][]{Ting2012, Ness2022, Weinberg2022, Griffith2024}, although see \citep{Ting2022, Manea2023}. The Milky Way halo, which shows more abundance diversity than the disk, offers insight into different mechanisms, epochs, and environments of star formation. This diversity in the halo is likely driven by the many individual systems, with different masses and star formation histories, that have built up the halo \cite[e.g.,][]{Naidu2020, Horta2023} and the stochastic enrichment at early times \citep[e.g.,][]{Ji2015}. 

Despite being embedded in an era of large data volumes, the mechanisms governing the origin of the $\alpha$-bimodality, the apparent low-dimensionality of the Milky Way disk, and detailed progenitor composition of the halo are disputed or weakly constrained. Simulations can provide insight into the holistic assembly of the Milky Way, and test which physical mechanisms drive observational characteristics. On the empirical front, there are growing opportunities to connect and compare Milky Way observations to other galaxies. This comparison enables a contextualisation of the attributes of the Milky Way. 

Nearby galaxies can be studied using samples of individual stars. In particular, bright stellar tracers such as long period variables, tip of the red giant branch stars, and planetary nebulae are useful tools \citep[see][]{Parto2023, Ren2021,  Hartke2022}. The iron and $\alpha$-element abundances as well as radial velocities of stars in a large number of dwarf galaxies has been catalogued via medium-resolution spectroscopy (R=6,500) \citep[e.g.][]{Kirby2009, Kirby2010, Kirby2011, Escala2020, Longeard2020, Walker2023}. High-resolution  (R $>$ 25,000) observations of small samples of stars in nearby dwarf galaxies, observed with 8-m class telescopes, have provided benchmark, high-fidelity samples from which individual abundances have been determined \citep[see][and references therein]{Venn2008, Tolstoy2009, S2019, J2023, Shetrone2003, Aoki2020}.  Large spectroscopic surveys like \apogee\ have also recently targeted bright giants in the local group. This has provided a set of  dwarf galaxy individual element abundances homogeneously measured from medium-resolution (R=22,500), high SNR ($>$ 70) spectra \citep{Hasselquist2021}. The abundance measurements have been derived with the ASPCAP pipeline \citep{GP2015}, and are therefore on an element abundance scale that is consistent with hundreds of thousands of stars observed by \apogee\ in the Milky Way. The diversity of distributions of individual element abundances in other galaxies indicates that they have experienced different enrichment and star formation histories. In dwarf galaxies in particular, the  different abundance characteristics compared to the Milky Way are explained by mass-dependent mixing and feedback \citep{AE2018, AE2019, Kobayashi2020b}. The systematic differences between the Milky Way and dwarf galaxy abundances may also be associated with the loss of the circumgalactic medium in dwarf galaxies \citep{Zhu2024}.

In the extra-galactic realm, integrated spectroscopy and resolved photometry of the local, and increasingly, high-redshift universe has unveiled diversity within and between stellar systems. Surveys like Physics at High Angular resolution in Nearby GalaxieS \citep[PHANGS;][]{phangs2019, phangsmuse2022}, Local Volume Mapper \citep[LVM;][]{LVM} and Generalising Edge-on galaxies and their Chemical bimodalities, Kinematics, and Outflows out to Solar environments \citep[GECKOS;][]{Geckos} are providing spatially resolved spectroscopic data of nearby systems. Instruments including The Multi-Unit Spectroscopic Explorer (MUSE), the Hubble Space Telescope (HST), and the James Webb Space Telescope (JWST) provide the data to connect star formation histories to galactic structure and kinematics, and study environments and galaxy assembly over time \citep[see][and references therein]{DW2023, Williams2021, Gadotti2019, Milone2023, JF2024, Martig2021, Conroy2019, DW2014, Bell2005}. Next generation infrastructure like the Multi-AO Imaging Camera for Deep Observations \citep[MICADO;][]{MICADO} and the MCAO Assisted Visible Imager and Spectrograph \citep[MAVIS;][]{MAVIS,MAVIS2} will resolve individual stars and provide a deeper and higher resolution mapping of abundances, as well as spatial and kinematic properties of galaxies to higher redshift.

As the numbers of spectra have risen, there has been a corresponding expansion in the utility of so-called data-driven methods. Specifically, whereby subsets of observational data are used to build models to learn about larger samples of observational data, or else to propagate information between surveys \citep[e.g.,][]{Ness2015, Casey2016, Leung2019, Feeser2022, Rice2020, Rains2024, Laroche2024, Thomas2024}. Data-driven methods have found notable success in labeling low signal to noise and low resolution data with precise stellar parameters as well as abundances \citep[e.g.,][]{ Ho2017b, Ho2017a, Wheeler2020, Zhang2024, Andrae2023}.

The large volumes of data and new techniques are providing the framework to realise the Milky Way in a `cosmological context', whereby we can directly compare the distribution of element abundances and morphological features of the Milky Way to other galaxies. A prohibiting factor in doing this, however, is the need for a consistent comparison of the extra-galactic and Galactic observations and subsequent measurements \citep[e.g.][]{Martig2021}. One approach to homogeneously compare and contrast the Milky Way and other galaxies is to reconstruct the Milky Way as an integrated galaxy itself \citep[e.g.][]{Lian2023}. Another is via large multi-object spectroscopic surveys that have targeted stellar populations outside the Milky Way in addition to Milky Way targets. 

We combine the benefits of the substantial growth in the volume of extra-galactic Integral Field Unit (IFU) data, Milky Way stellar data of different qualities, and the demonstrated utility of data-driven methods of the last decade. The spectroscopic data that exists over a large range of quality, resolution and spatial scales provides an opportunity to connect and compare the properties of the Milky Way to other galaxies in new ways.  This opportunity was recently recognised by \citet{Wang2022}, who showed that data-driven methods are effective in deriving information, including individual abundances, from low resolution resolved MUSE IFU spectra, using a model built using Milky Way spectra.

We introduce the first paper in our ``extra-galactic ArChAeology with Cannon Inferred Abundances from resolved integral field unit Spectra''  \textsl{(ACACIAS}) program. The goal of \textsl{ACACIAS} is to derive and subsequently compare the abundance characteristics of the Milky Way and other galaxies in the nearby group using low fidelity data (Resolution, R=1800, Signal to Noise Ratio, SNR$>$ 10). There is diversity among the local group systems in mass, morphology and star formation properties.  This homogeneous census will serve to elucidate connections between element abundance distributions as a function of galactic properties. Our approach takes advantage of the demonstrated ability of data-driven models to label low-SNR, low resolution spectra with stellar parameters and abundances. We subsequently use nomenclature that is indicative of the nature of our abundances as inferred labels from a model that uses the entire stellar spectrum (and not abundance measurements from isolated individual element lines which are typically inaccessible at low resolution for most elements). The abundances that are derived via this approach are directly on the ``scale'' of the \apogee\ survey, enabling a direct comparison with the Milky Way, as well as a series of dwarf galaxies observed for that program. 

We introduce the method, and showcase the results for a first galaxy in this series, NGC~6822, for which individual stars have been resolved from a MUSE IFU data-cube. To infer stellar parameters and abundance labels we build a model trained on R=1800 \lamost\ survey spectra \citep{Lamost2012}, where we have taken high quality stellar parameter and abundance labels from \apogee\ \texttt{DR17} \citep{Majewski2017, Ab2022}.  NGC~6822 is an isolated and relatively large dwarf irregular barred galaxy ($\sim 10^8$ M$\cdot$) with evidence for an enhanced star formation rate over the past 500 Myr \citep{Fusco2014}. The \Gaia\ proper motions suggest that it may have obtained its twisted halo morphology via an encounter with the Milky Way within the Milky Way virial radius $\sim$3-4 Gyr ago and even lost some stellar constituents to the Milky Way \citep{Zhang2021}. There are also claims that NGC~6822 is a `polar
ring' galaxy based on kinematic misalignment of a sample of C-stars relative to the neutral hydrogen gas \citep{Demers2006}.  Additionally,
old extended globular clusters have been discovered well out into the
halo, with alignment (in projection), suggesting possible accretion
origin \citep{Hwang2011, Huxor2013}.

This paper is organised as follows: Section 2 outlines our data and methods, Section 3 presents our results, in Section 4 we discuss our results and in Section 5 we summarise the work.

\section{Data and Methods} 

\subsection{NGC~6822 observations}

Observations of NGC\,6822 were taken as part of MUSE Science Verification programs on 2014 August 24 and 25. A total of 4 $\times$ 1320s exposures were obtained in a single pointing focused southeast of the central bar region, corresponding to ``Grid 4'' from \citet{Cannon2012}. At the distance of NGC\,6822 \citep[$459\pm17$~kpc;][]{Mcconnachie2012}, the $\sim1^{\prime}\times1^{\prime}$ MUSE field subtends roughly 130~pc. Each exposure was rotated by 90 degrees relative to the previous in order to average over the MUSE slicer pattern in the combined data cube.  Pairs of object exposures were taken along with a short 120s sky exposure in an object-sky-object pattern. 

The data were reduced using v2.8.5 of the ESO MUSE instrument pipeline \citep{Weilbacher2020} controlled via \pmp\ \footnote{https://github.com/emsellem/pymusepipe}. This includes bias and overscan subtraction, flat fielding, wavelength calibration, telluric correction, and sky subtraction. We align the individual exposures in \emph{x} and \emph{y} via cross-correlation of their white-light images, using the first exposure as a reference. We then identify individual objects in each of the MUSE exposures and use these to fine tune the alignment in both spatial position and rotation using a simple grid search over rotational offset, $\Delta\theta_\mathrm{rot}$, where at each step we re-estimate the best \emph{x} and \emph{y} offsets via cross correlation and compute the r.m.s. deviation of object positions; in most cases the derived shifts are small. Reconstructed cubes are then combined using a simple mean. The PSF FWHM measured from bright stars in the combined data ranges between 0\farcs84 at the blue end to 0\farcs68 in the red, with an average of 0\farcs75.

%final6
%run readin_cutout_ryan.py 
\begin{figure}
\centering
\includegraphics[width=\columnwidth]{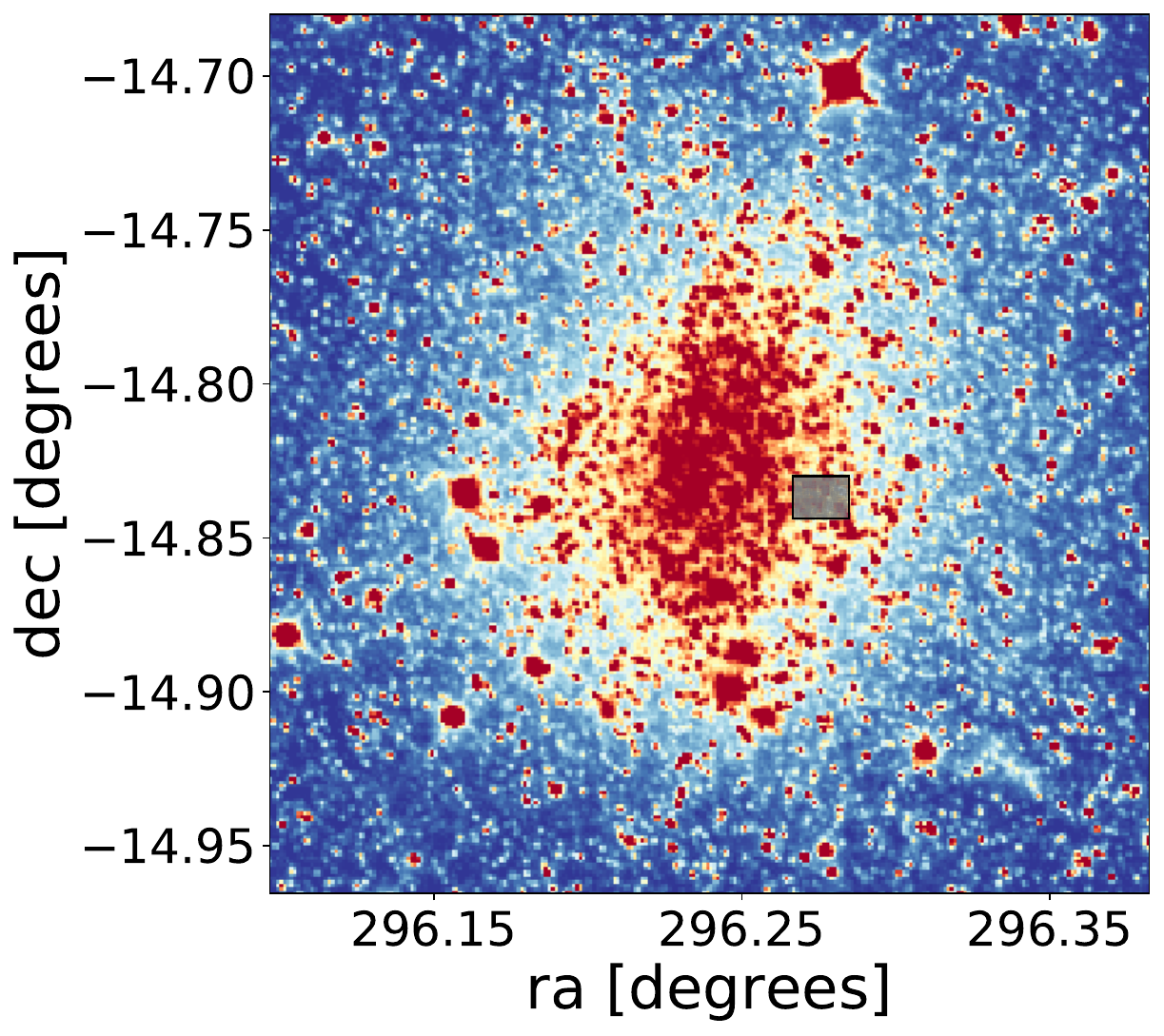}
\caption{An image of NGC~6822 taken from the Legacy Sky Survey, indicating in the shaded grey region the locations of the 1716 resolved spectra extracted from the footprint of the MUSE IFU field of view.}
\label{fig:one}
\end{figure}

\subsection{Resolved Spectra Extraction from MUSE Data}

We extract spectra for individual stars using \pampelmuse\, which is specifically designed for the extraction of stellar spectra in crowded fields. Our use of \pampelmuse\ here follows closely the methods described by \citet{Kamann2013} and \citet{Kamann2018}, which we summarize below.

In addition to the MUSE data cube for NGC\,6822, \pampelmuse\ requires as input a reference catalogue of source positions that serve as the basis for its spectral extraction routines. We use here the HST/ACS imaging catalogues for Grid 4 from \citet{Cannon2012}, which also served as the basis for our initial field selection. We use both the source positions and magnitudes in the F814W band to define objects in the MUSE footprint, since the majority of the band pass falls within the MUSE spectral range (c.f. the bluer F475W filter). We add to these catalogues eight additional (bright) sources that were saturated in the original ACS imaging. 

We use \pampelmuse\ to generate an initial set of fitted parameters—namely $\beta$ and FWHM of the Moffat Point Spread Function (PSF) and source positions as a function of wavelength—using data binned by 50 layers in the spectral direction. There are $\sim$17,000 sources in the input catalogue that fall within the MUSE footprint, with $\sim$6000 of those above the 50\% completeness limit of $m_\mathrm{F814W} = 24.9$ mag quoted by \citet{Cannon2012}. The location of these data with respect to Legacy survey photometry\footnote{$https://www.legacysurvey.org/$} is shown in Figure \ref{fig:one}. This corresponds to a radial distance of $\sim$ 18kpc along the disk. 

We identify 1716 sources as nominally ``resolved'' in this initial pass over the data, and it is these stars that have their spectra extracted in subsequent steps. We include the remaining sources as an unresolved background component when extracting the resolved source spectra. We model the change in source position as a function of wavelength using a 5$^{\mathrm{th}}$ order polynomial, but verified that the final extracted spectra do not change substantially if we adopt a 3$^{\mathrm{rd}}$ or 7$^{\mathrm{th}}$ order polynomial.

The final step in our processing of the extracted MUSE spectra is to degrade and resample the data to match the LAMOST spectra. We assume the wavelength-dependent resolution curve for MUSE from \citet{Bacon2017}, and smooth the spectra to a constant $R=1800$. Data are then logarithmically rebinned to a fixed pixel size of 69~km\,s$^{-1}$. Finally, we continuum normalise the spectra by dividing it by a Gaussian smoothed version of itself, using a kernel with a standard deviation of 50\AA{} truncated at $\pm$150\AA{} from the line center as done for \lamost\ spectra in  \citet{Wheeler2020}.

For each spectrum we output a wavelength array, flux array and flux uncertainty array. We use the ratio of the flux and flux uncertainty arrays to calculate the signal to noise ratio (SNR) as a function of wavelength for each spectra, and calculate the median of these values to serve as an SNR measure for each star.  

\begin{figure}
\centering
\includegraphics[scale=0.3]{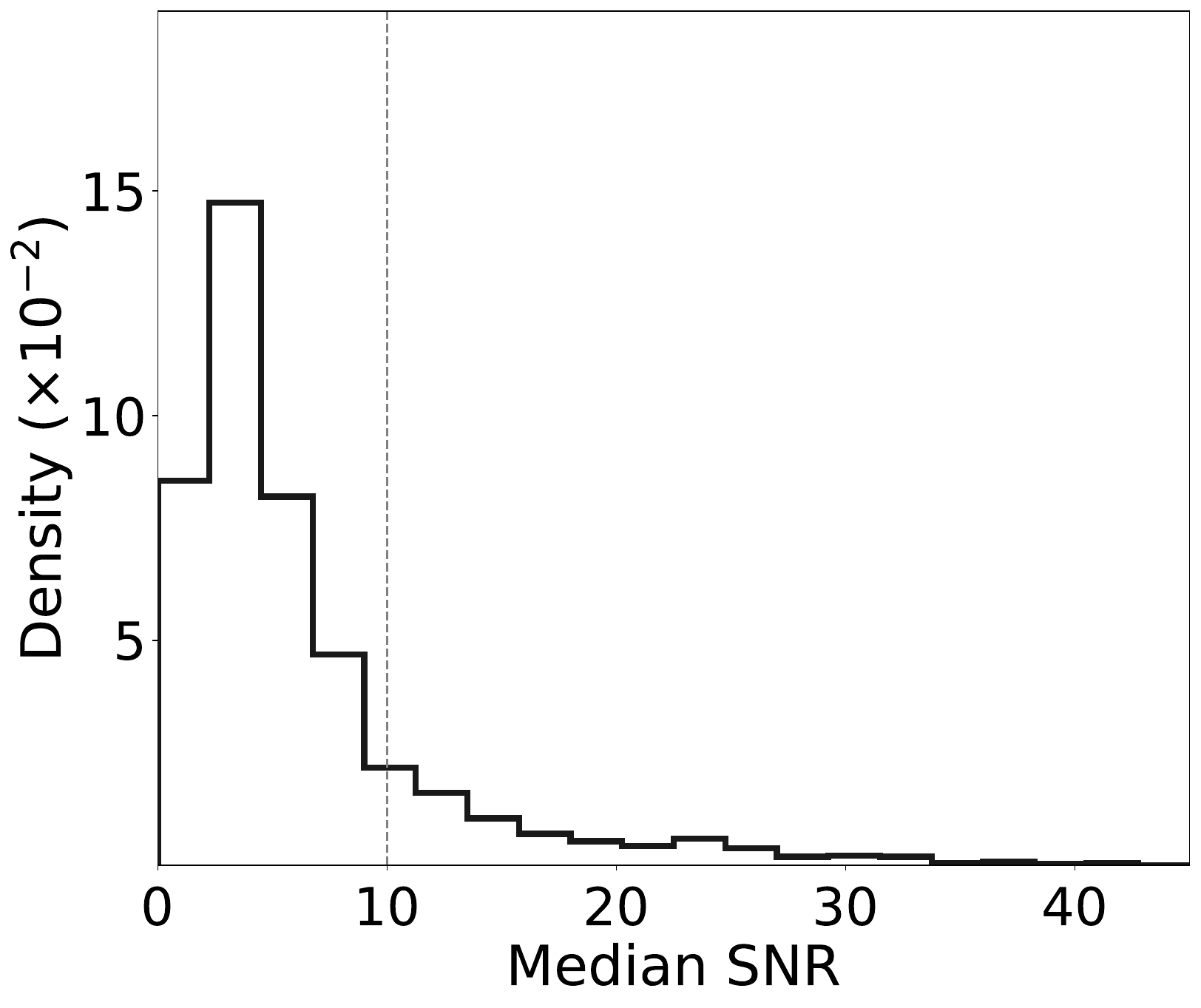}
\caption{The distribution of the median SNR for all 1661 resolved stars; a subset of the 273 with SNR $>$ 10 are used to report the element abundances}
\label{fig:two}
\end{figure}

We have a total of 1716 spectra, and of these 55 are poor extractions that have brighter neighbours within 3 pixels. These occur almost exclusively at the side of the field with the higher background stellar density.  The Signal to Noise Ratio (SNR) distribution of the remaining 1661 spectra is shown in Figure \ref{fig:two}. Note that in detail, over the large wavelength range of the observations, the SNR changes by a factor of around $\sim$2 for each spectra. However, the median serves as a useful number to serve as a basis of reference and comparison. 

%final6
%run -i makevelsnrhist.py 

We ultimately utilise only the stars with SNR $>$ 10 for our analysis, and these are indicated in the tail of the distribution to the right of the dashed vertical line shown in Figure \ref{fig:two}. The lower SNR spectra can likely be stacked and used in future work, by combining stars with the same inferred stellar parameters. However, this is not necessary for the purposes of our analysis, as we have sufficient numbers with this lower limit cutoff to proceed with analysis and to obtain results regarding the abundance distribution of NGC~6822 using the relatively higher quality subset of spectra available.

\subsection{Radial velocity correction}

%final7
% run -i makevelhist.py 
\begin{figure}
\centering
\includegraphics[scale=0.3]{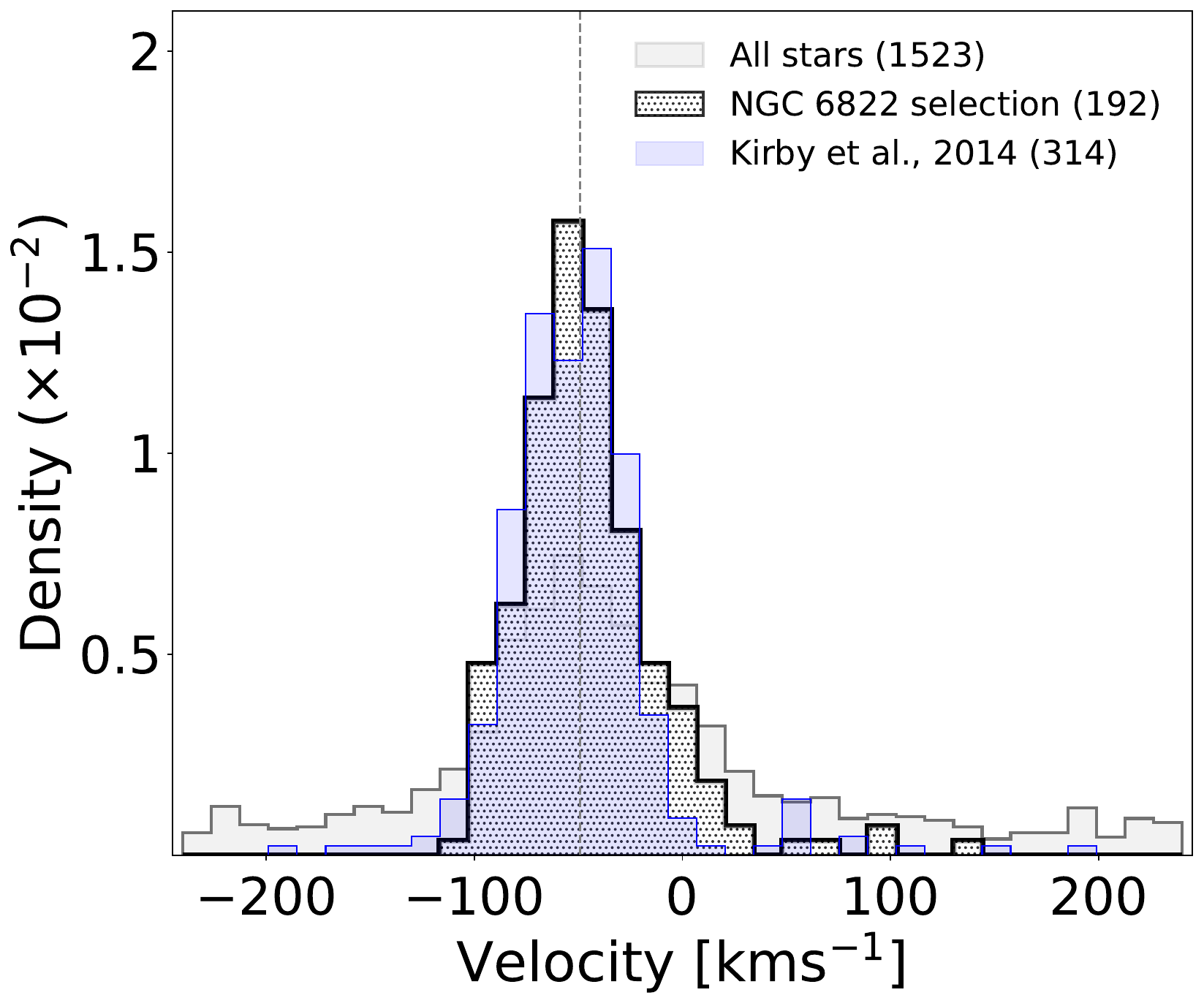}
\caption{The velocity distribution for 1523 resolved stars in the NGC~6822 field for which we report velocities (we exclude the 8 percent of stars with grid edge velocity solutions) shown in the light grey histogram. The dot hatch filled histogram shows the distribution for the 179  stars with SNR $>$ 10 that meet criterion for membership described in Section 2.3. The median velocity ($-49.9$kms$^{-1}$) is indicated with a dashed vertical line. The radial velocities for the 314 NGC~6822 stars reported in \citet{Kirby2014} are also included for comparison in the light blue filled histogram; there is good agreement in the mean velocity and overall velocity distribution between the studies.} 
\label{fig:three}
\end{figure}

We cross-correlate each of the 1661 spectra with the 0$^{th}$ coefficient of the model that we employ to determine the stellar parameters and individual abundances, which is described in Section 2.4. We use the Python function \textit{crosscorRV}, in doppler mode, stepping over a velocity interval of V$_{range}$ = $\pm$250 km s$^{-1}$. This function re-bins the observational data onto the same wavelength frame as the model and finds the cross correlation function; the peak of a fifth order polynomial fit to the cross correlation function output represents the velocity of the star. We exclude the $\sim$8 percent of stars with velocities at the grid edge. The histogram of the radial velocity distribution of the observations for the remaining 1523 stars is shown in Figure \ref{fig:two}. The grey shaded histogram shows the stars with SNR $>$ 10 only that meet our NGC~6822 membership cuts described in Section \ref{sec:membership}. The radial velocity distribution for the set of stars has a mean of $V_{bary} = $$-46.8 \pm$ 32 kms$^{-1}$ measured with a confidence of $\pm$ 2.3kms$^{-1}$ (and a median of $-49.9$ kms$^{-1})$. This agrees well with previous estimates \citep{Thompson2016, Kirby2013}.

%final5
%crosscor4_resfix_velfix_errordraw_norebin_velonly.py
%run -i runmeanerrval.py
\begin{figure}
\centering
\includegraphics[scale=0.3]{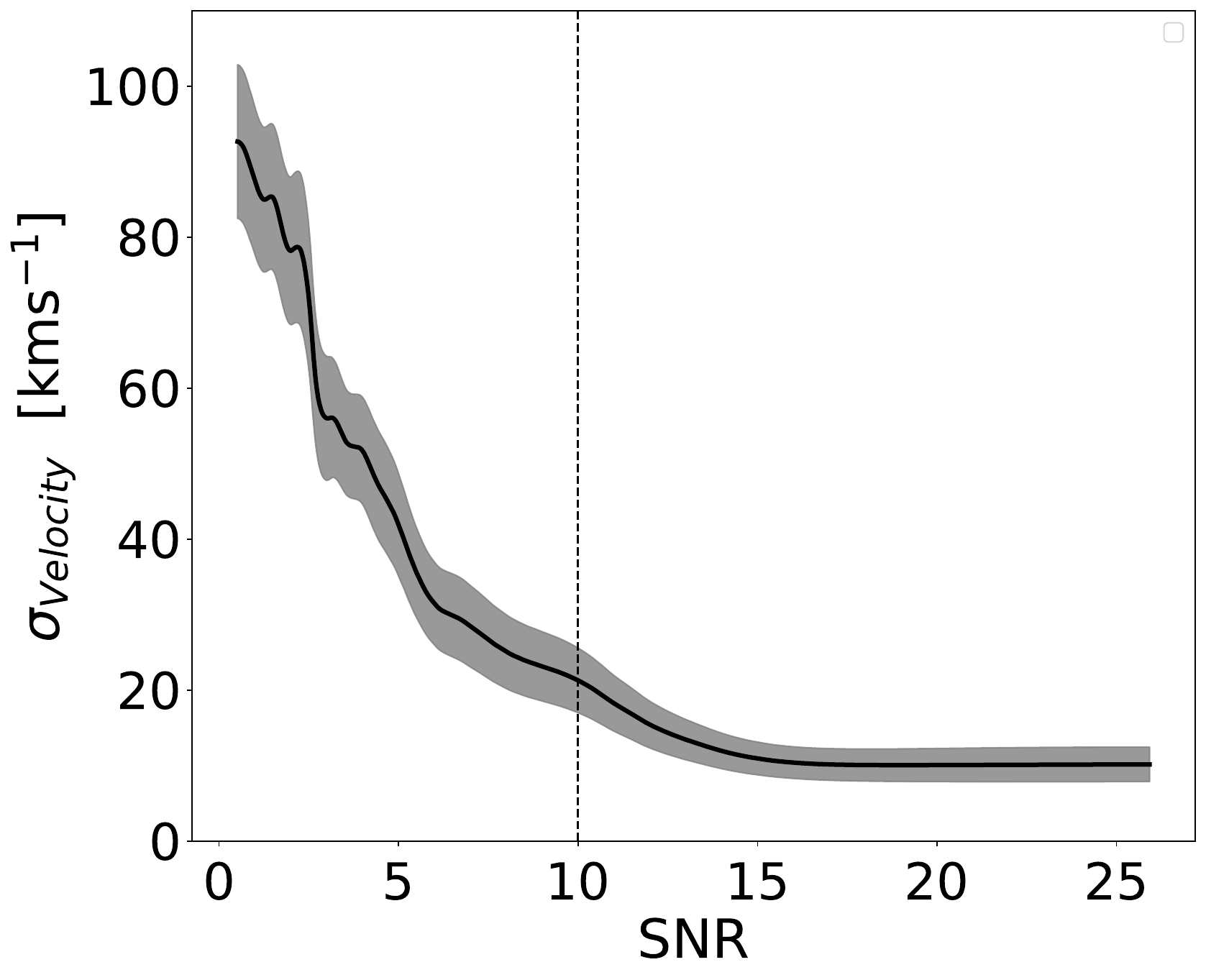}
\caption{The velocity uncertainty for 1523 resolved stars in NGC~6922 as a function of SNR. The vertical line at SNR = 10 indicates the threshold above which we report stellar labels inferred from the spectra. The uncertainty across SNR = $10 - 20$ is $\sigma_{velocity}$ = $\approx$ 21 - 10~kms$^{-1}$.}
\label{fig:velprecision}
\end{figure}

Our radial velocity measurement uncertainty is subsequently determined by resampling each observation from its noise distribution, at each wavelength, performed 10 times. The uncertainty on the radial velocity measurement as a function of SNR is shown in Figure \ref{fig:velprecision}. 
The precision floor is similar to that found in \citet{Kamann2020} for resolved spectra from MUSE. 

\subsection{Photometry and NGC~6822 Membership} \label{sec:membership}

%final7/newmodel/revc/makecolormag_multi_velonly.py 
\begin{figure}
\centering
\includegraphics[scale=0.4]{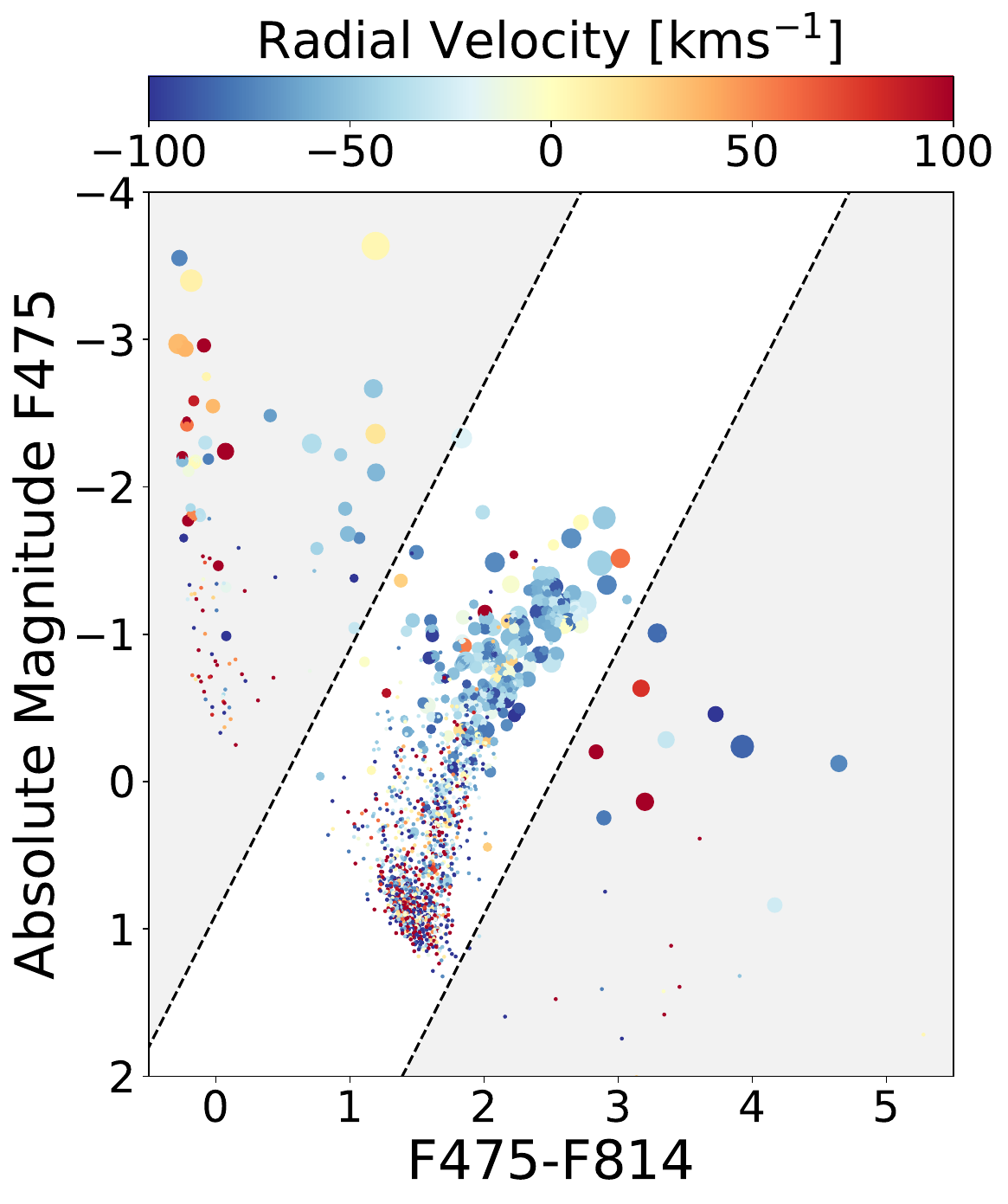}
\caption{The colour-magnitude distribution of the 1523 resolved stars, assuming a distance to NGC~6822 of 500~kpc. The sources are coloured by their determined radial velocity. For spectra with SNR $>$ 10, the size of the points is scaled by SNR, with larger points corresponding to higher SNR. The spectra with SNR $<$ 10 are shown in the smaller size circles. The un-shaded region of the figure represent the red giant branch region of NGC~6822, and the shaded regions are presumably Milky Way Galaxy foreground stars which have high positive radial velocities, in contrast with the mean velocity of the NGC~6822 system. }
\label{fig:velcmd}
\end{figure}

In Figure \ref{fig:velcmd} we show the colour-magnitude diagram for the NGC~6822 field sources using the HST/ACS F475W and F814W filters, with the absolute magnitude in F475W computed assuming a distance to NGC~6822 of 500~kpc. The white un-shaded regions represent the red giant branch region of NGC~6822. The grey shaded regions are presumably Milky Way contaminants, or hotter stars far outside of the reference objects \teff $>$ 6000K. The radial velocity is shown in the colour of the points and the point size is scaled by the SNR, with larger points corresponding to the brighter higher SNR observations (integration time is the same for all objects). The points that are off the red giant branch are correspondingly at preferentially positive radial velocities.

%final7/newmodel/revc/maketraining.py
\begin{figure*}
\centering
\includegraphics[scale=0.5]{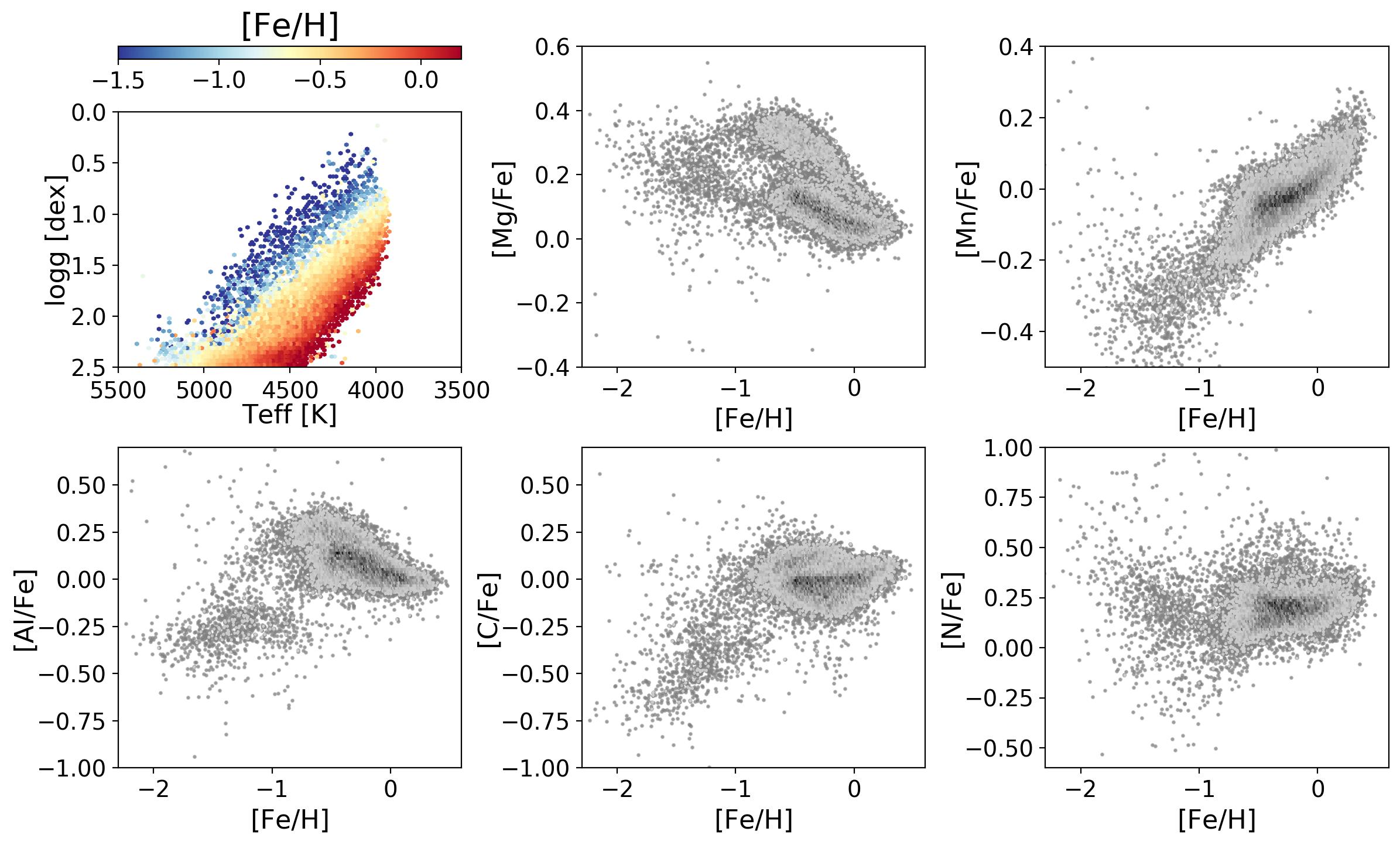}
\caption{The \apogee\ label distributions of the 19,296 reference objects from the \apogee\ $\times$ \lamost\ set of stars that are used for training \tc's model, as described in the text.}
\label{fig:reference}
\end{figure*}

\subsection{Stellar Parameter and Abundance Determination with The Cannon}

To determine stellar parameters and abundances for the resolved NGC~6822 spectra, we use a red giant model built using 19,296 training stars that are observed by both \apogee\ and \lamost. The spectra that are used to train the model are from \lamost\ (R=1800) and the labels are from \apogee\ (R=22,500). We select the following ten labels from \apogee\ DR17 \citep{Ab2022} to build the model. These labels are:  \teff, \logg, $v_{\mbox{mic}}$, A$_k$, [Fe/H], [Mg/Fe], [Mn/Fe], [Al/Fe], [C/Fe], [N/Fe]. The training stars cover the expanse of the \apogee\ abundance space and are shown in Figure \ref{fig:reference}. They have the following properties: \\

\begin{footnotesize}
\noindent
$\mathrm{SNR}_{\lamost}$ $>$ 100 \texttt{OR} if logg $<$ 1, $\mathrm{SNR}_{\lamost}$ $>$ 50 \\
$\mathrm{[X/Fe]}_{err}$ $<$ 0.2 \\
3986 $<$ \teff $<$ 5640~K \\
0.32 $<$ \logg $<$ 2.5~dex \\
--2.15 $<$ [Fe/H] $<$ 0.48 \\
--0.32 $<$ [Mg/Fe] $<$ 0.49 \\
0 $<$ A$_k$ $<$ 0.4 \\
0.3 $<$ $v_{\mbox{mic}}$ $<$ 3.5 \\
--1.25 $<$ [Mn/Fe] $<$ 0.28 \\
--0.82 $<$ [Al/Fe] $<$ 0.72 \\
--1.0 $<$ [C/Fe] $<$ 0.91 \\
--0.52 $<$ [N/Fe] $<$ 1.36 \\
\end{footnotesize}

The model training and validation procedure is similar to the implementations outlined in \citet{Wheeler2021}, for a \lamost\ $\times$ \galah\ model, and \citep{Ho2017a,Ho2017b},  for a \lamost\ $\times$ \apogee\ model. For this work we select a \lamost \ $\times$ \apogee\ training set so as to reach stars with \logg\ $<$ 1, at the tip of the red giant branch. This reaches the extreme of the parameter space of the faint dwarf galaxy observations.  There are a generous number of 213 stars in the training set with \logg\ $<$ 1.0 dex. As \tc\ employs a simple quadratic function and effectively acts as a means of interpolation between spectra that are typically well described by this model form, very few objects are required for training. A more important feature of the reference objects than dense sampling is their parameter space coverage \citep{Ness2015, Walsen2024, Fabbro2018}. Importantly, the generative nature of \tc\ means that a model spectra can be created for every observation given the inferred labels. This can be used to test, via a $\chi^2$ comparison with the observed spectra, if the labels subsequently do describe the spectra.

The model formalism is, as previously, a second-order polynomial description of the flux, given the set of labels. The element labels included here cover a number of nucleosynthetic channels (Al; odd-z, C, N; light, Fe, Mn; iron-peak, Mg; $\alpha$). These channels are associated with primary production in Supernovae Type Ia (SN~Ia) (for Fe, Mn), core collapse supernovae which are predominantly Supernovae Type II (SN~II) (for Mg, Al), and Asymptotic Giant Branch (AGB) s-process element production (C,N).

\begin{figure*}
\centering
\includegraphics[scale=0.55]{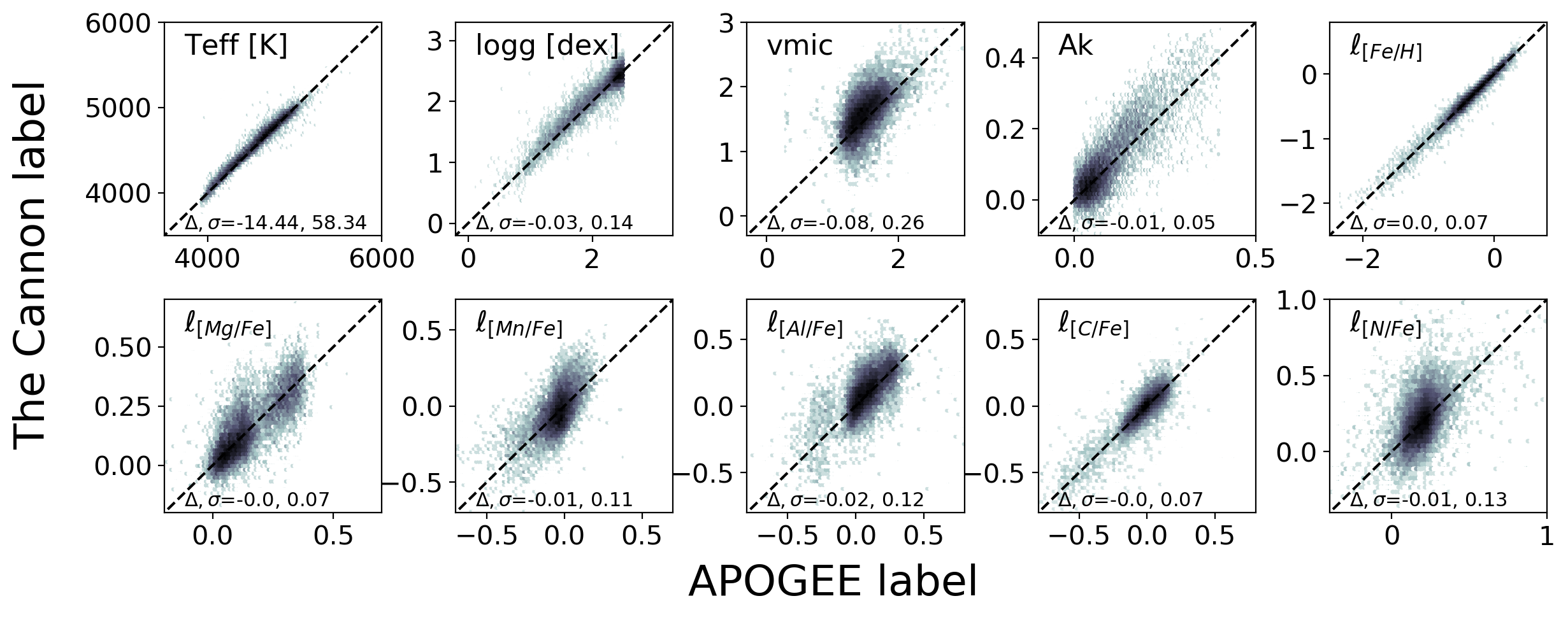}
\caption{Cross-validation of the model build using 19,296 reference objects using \lamost\ spectra and \apogee\ labels. This figure represents a take-25 percent out test, whereby one quarter of the reference objects are removed iteratively and the inference reported for the held-out stars. The metrics reported are bias and rms scatter around the 1:1 line with the poorly fit ten percent of stars with $\chi^2$ $>$ 15,000 removed.}
\label{fig:xval}
\end{figure*}

We use cross-validation to test the model performance. Figure \ref{fig:xval} shows the measured versus predicted labels for a take-25 percent out test; whereby one quarter of stars are removed from the model, the model trained and tested on the left out stars, four times. Each sub-panel reports the bias and standard deviation of the (predicted-reference label) distributions. The element abundance labels are predicted to within $<$ 0.15 dex (on average across the full parameter space of the reference objects).

Note that the model may be able to accurately extrapolate in a limited region outside of the boundaries of the reference objects, since the bounds from cross-validation are primarily driven by the extent of the training set, rather than observed bias.

%newmodel/spectra_obsidx2/plotfitspec.py

At training time, the model relates stellar flux to already known stellar labels given the reference objects and is solved independently at each wavelength. This generates a set of coefficients; a 0$^{th}$ term, linear terms, cross terms and squared terms, as well as a scatter term which parameterises the goodness of fit of the model at each wavelength. This model is subsequently deployed on new stellar spectrum, at test time, to infer stellar labels given an entire spectrum. This generates the best fit labels and their uncertainty, under a maximum likelihood formalism. Given the model coefficients, these labels can also be used to generate the corresponding best-fit stellar spectrum to compare to the observational data to assess the goodness of fit of the model spectrum. 

%final7/newmodel/revc/makemodelspectrum2_movecbar2.py 
%makecmap.py 
\begin{figure*}
\centering
\includegraphics[scale=0.58]{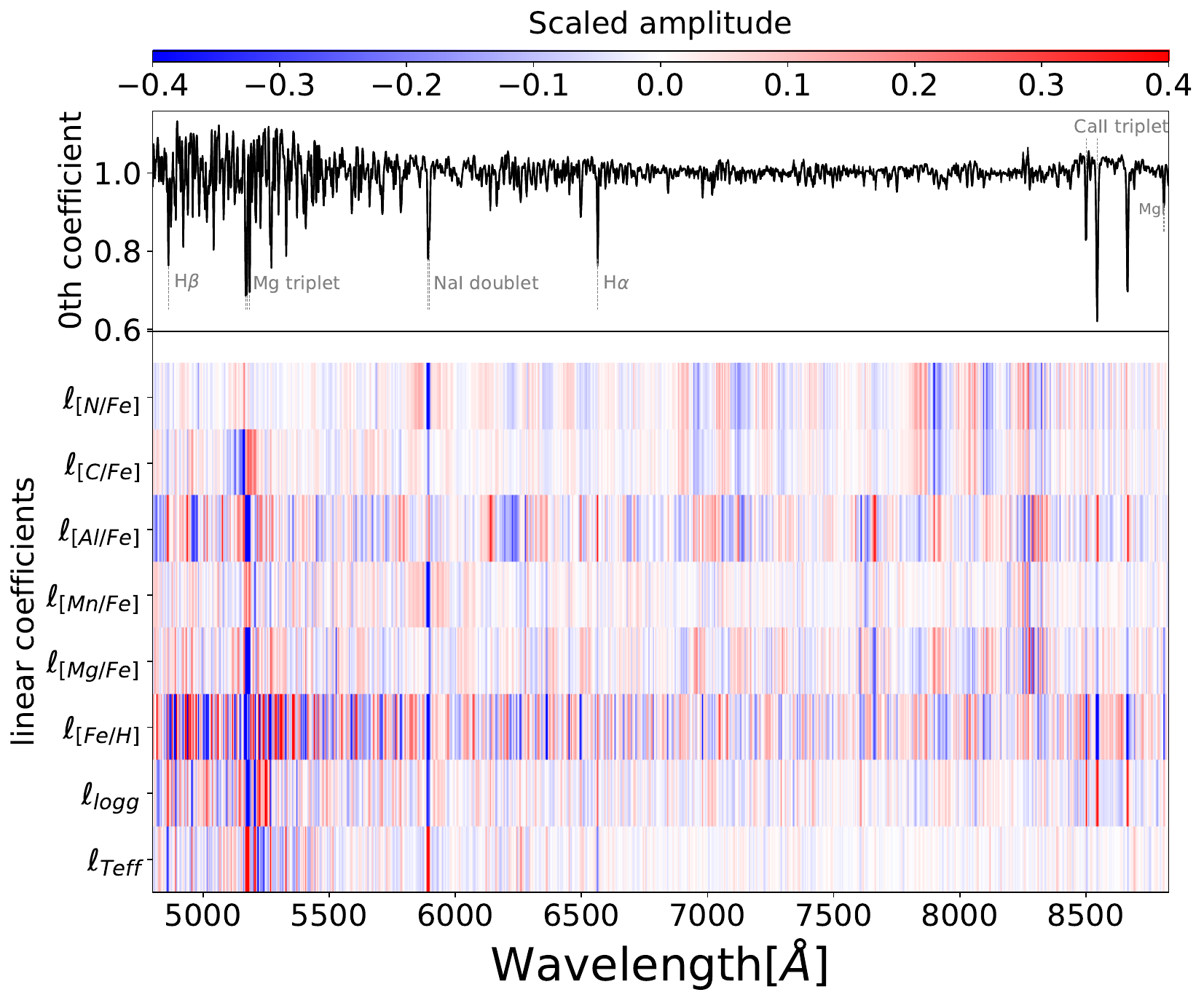}
\caption{A summary of a set of the coefficients of \tc's model that is used to determine the labels for the NGC~6822 spectra. The top panel shows the 0$^{th}$ coefficient of the model  (approximating the mean spectra of the reference objects) with some atomic line features indicated for reference. The matrix at bottom shows the first-order  coefficients of the model's labels. Each row is scaled to the absolute maximum value of that coefficient for improved visualisation. The first-order coefficient flux gradients are both positive and negative, shown in red and blue, respectively; white or lightly coloured regions indicate the flux is insensitive to the labels. }
\label{fig:model}
\end{figure*}

A subset of the model coefficients are shown in Figure \ref{fig:model}. The top panel shows the $0^{th}$ order coefficient (roughly the mean spectrum) and the matrix shows the wavelength on the x-axis and the scaled linear coefficient are shown in each of the 8 rows on the  y-axis. For improved visualisation the amplitudes of the coefficients shown are scaled by the absolute maximum value of each coefficient across the wavelength region. No coefficients across wavelength are identical. This is demonstrative that each label has in detail a different overall dependence on the flux.
The linear coefficients  visualised in Figure \ref{fig:model} demonstrate where the information comes from in the spectra pertaining to the labels; these are $\partial F / \partial \ell$ at the mean training set labels (where $ F$ is the flux and $\ell$ are the training labels). Higher amplitude regions of the coefficients mean that wavelength is more sensitive to the variation of the label in the reference objects. The MUSE IFU spectra do not cover the full model range and we use subsequently only the part of the model that covers the wavelength region of the MUSE IFU spectra. This is straightforward with \tc\ framework, as the coefficients are solved for independently at each wavelength.

At the inference step, the entire spectrum is used to determine the value of each label -- including for individual elements. This means that  information is likely some combination of intrinsic information, as well as element correlations. Element abundances are expected to vary the atmospheric structure and therefore the flux across wavelength in complex ways \citep[e.g.,][]{Ting2018O, Starkenburg2010, Battaglia2008, V2016, Gust2008, Muc2012}, which \tc, as well as other flexible fitting approaches, are able to capitalise on. This is particularly so at the top of the red giant branch where there are many molecules (blended features) that can be used to infer abundances via an analytical approach. Reassuringly,  the coefficients do have high amplitudes where the individual atomic features of the relevant lines are present \citep[see e.g.][]{Manea2023}. We test where \tc's model's features are associated with physical absorption lines relevant element labels in the Appendix Figure \ref{fig:model2}.

An assumption underlying responsibly deploying the data-driven approach to undertake element inference is that the training set of stars are representative of the test objects. Under this assumption, the relationship between stellar labels and stellar flux is subsequently consistent between both training and test sets of stars. The model can reasonably extrapolate in a limited range outside of the boundaries of the reference objects. Important in this context is the simplicity of the model and its analytical form, as well as the validation enabled by generating the theoretical spectrum to compare to the model spectrum.   

In summary, to prepare the spectrum for inferring labels with the model built using the \lamost\ spectra, the spectra has been (i) extracted from the datacube across the same wavelength grid and resolution as the \lamost\ spectral model at rest vacuum wavelengths (ii)  continuum normalised following the procedure used to build the model (iii) wavelength corrected for radial velocity doppler shift. The spectra is then ready for processing with \tc\ using the model built using \apogee\ labels and \lamost\ spectra for stars in common with the survey. The model's coefficients are utlised in concert with the flux and flux uncertainties of each spectrum to infer the labels corresponding to each spectrum, as well as to generate a best-fit model spectrum for those inferred labels.

%final7/newmodel/revc/makecolormag_multi_four.py  
\begin{table*}
\centering
\footnotesize
\begin{tabular}{|l l l l l l l l l l l l l l  |} 
 \hline
No. & SNR & ra [deg] & dec [deg] & \teff & \logg & vmic &  Ak &   $\ell_{[Fe/H]}$ & $\ell_{[Mg/Fe]}$ & $\ell_{[Mn/Fe]}$ & $\ell_{[Al/Fe]}$ & $\ell_{[C/Fe]}$ & $\ell_{[N/Fe]}$  \\ [0.5ex] 
 \hline\hline
50 & 15.5 & 296.277672 & -14.842069 & 4224.6 & 0.84 & 1.9 & 0.23 & -0.79 &  0.31 & -0.28 & 0.22 & -0.08 & -0.11\\
\hline
51 & 18.2 & 296.280388 & -14.833756 & 4315.3 & 2.23 & 0.59 & 0.08 & -0.64 &  -0.06 & 0.17 & -0.87 & -0.01 & -0.32\\
 \hline\hline
\end{tabular}
\caption{Excerpt from table that is available in full online for the 185  stars with SNR $>$ 10, for which we report stellar parameters and abundance labels. The online version of the table also includes the evaluated uncertainties on each label for each star. We use the nomenclature $\ell_X$ for abundances which indicates these are inferred labels from a model where they correspond to the following reference labels X: [Fe/H], [Mg/Fe], [Mn/Fe], [Al/Fe], [C/Fe], [N/Fe]. }
    \label{tab:results}
\end{table*}

\section{Results}

%run -i compileerrors.py
%final7/newmodel/revc/runmeanerr_all.py 
\begin{figure}
\centering
\includegraphics[scale=0.35]{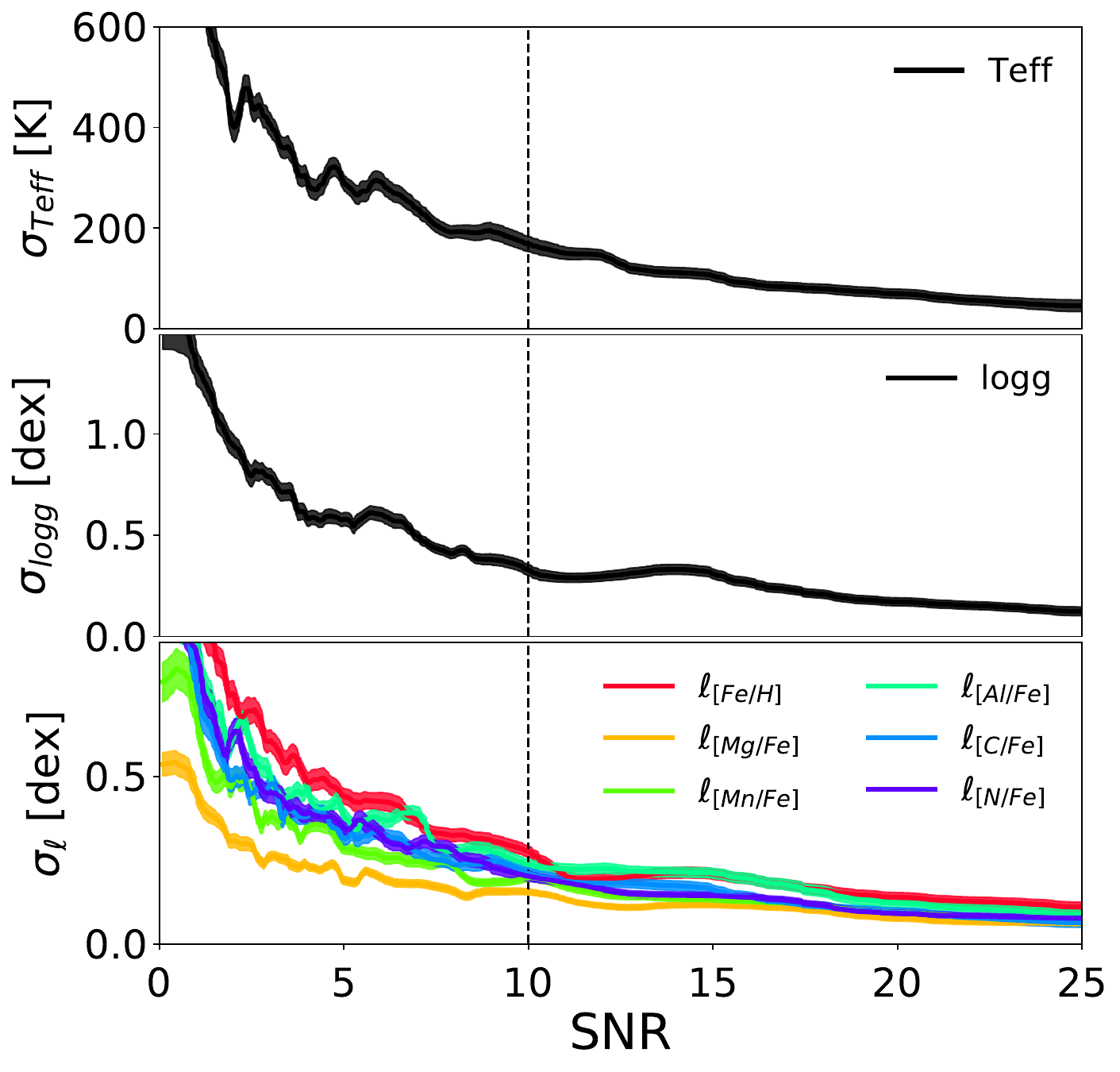}
\caption{The uncertainty of the inferred abundance labels as a function of SNR. The labels for which the uncertainty is mapped are indicated in each subplot and the colour-scale gives a range of uncertainty values. Only abundances for stars above SNR $>$ 10, indicated with the dashed line, are considered for analysis and reported.}
\label{fig:precision}
\end{figure}

\subsection{Labels inferred using \tc}

We infer the following labels for each resolved stellar spectra: \teff, \logg, $v_\mathrm{mic}$, X = \feh, \mgfe,  [Mn/Fe], [Al/Fe], [C/Fe], [N/Fe]. We chose to report all inferred abundances with the nomenclature $\ell_X$ to emphasise these are derived using the entire spectrum as per the distribution of label information learned at training time. 
While this deviates from the conventional approach of using individual lines pertaining to the element being measured, this has been shown to be both precise and accurate within the boundaries of the reference object labels \citep[e.g.][]{Ness2015, Casey2016, Buder2018, Manea2023}. In Sections 3.2-3.4, we will show that these labels have amplitudes that are consistent with expectations for dwarf galaxies and prior studies.

Figure \ref{fig:precision} shows the uncertainty of the inferred labels for objects with  as a function of SNR for the 1523 resolved sources. The uncertainty is determined by a Monte-Carlo procedure, repeating the label inference using a new realisation of each spectrum drawn from the noise distribution at each wavelength, ten times. The uncertainty on each label is calculated as the $1-\sigma$ standard deviation of the ten iterations.  As expected, the uncertainty increases substantially at low SNR. At SNR $=$ 10 the median uncertainty on \logg\ $=$ 0.35, on \teff\ $=$ 170K and the element labels range between $0.15-0.28$. We use the median uncertainties as a reference of the model performance but there is a large uncertainty range on a per star basis. We discuss and show individual abundance label uncertainties in more detail in the Appendix, Figure \ref{fig:precision2}.

%final7/newmodel/revc/makecolormag_multi_four.py
\begin{figure*}
\centering
\includegraphics[scale=0.40]{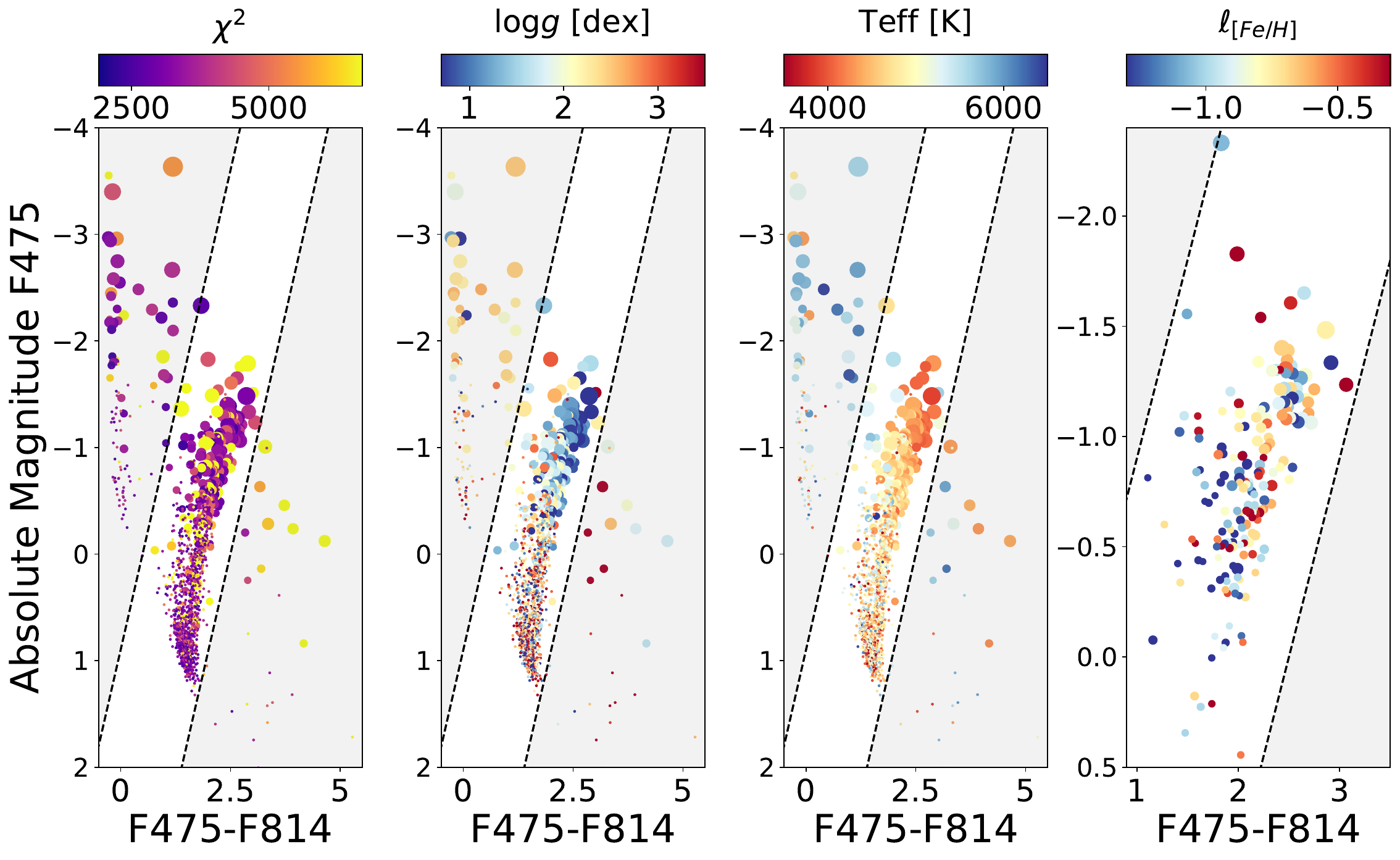}
\caption{The colour-magnitude plane for 1523 stars in the first three sub-panels, from left to right, coloured by $\chi^2$, \logg, and \teff, respectively. Similarly, to Figure 4, these assume a distance of 500~kpc to the cluster; the un-shaded region inside the lines indicates the stars considered for analysis as red giant members of NGC~6822. Note that the \logg\ of non-members are consistent with being main sequence stars, as well as stars at the base of the red giant branch. The panel at far right shows a zoom in of the 185  stars that meet quality cuts for which the individual abundance labels are reported, of $\chi^2$ $<$ 7947, logg $<$ 3.25 and SNR $>$ 10. The points are scaled by SNR with highest SNR being the largest symbols and stars with SNR $<$ 10 all having the same small marker size. A few percent of stars (6 of 192) show unphysical labels for their parameter space. These are possibly mislabeled main sequence stars but do not impact the results significantly and are not excluded.}
\label{fig:cmdall}
\end{figure*}

Figure \ref{fig:cmdall} shows the colour magnitude diagram as introduced earlier in Figure \ref{fig:velcmd}. Figure \ref{fig:cmdall} is now coloured by the stellar parameters inferred by the model as well as the $\chi^2$ of the data-model spectrum for each object. At far left, the $\chi^2$ distribution shows that most of the objects are distributed around $\chi^2$ $\approx$2000-4000, with a few outliers. The \logg\ distribution in the second panel from left shows that the red giant branch behaves as expected, with the lowest \logg\ values present the top of the red giant branch. The grey-shaded Milky Way ``contamination regions" of the colour-magnitude diagram are regions of inferred  high \logg\ stars, which are the likely foreground contamination main sequence stars. The \teff\ label similarly behaves as expected, with the coolest stars at the top of the red giant branch and hotter stars with smaller HST filter colour values.

We implement a series of quality cuts to limit our sample to where we have confidence in our results as members of NGC~6822 with robustly labeled spectra. These are as follows: (i) to be along the red giant branch only within the white un-shaded region of the colour magnitude diagram (ii) to exclude main sequence stars, admitting  stars with \logg\ $<$ 3.25 only (iii) to ensure a good generated model fit to the data and admit only stars with $\chi^2$ $<$ 7947 which is 3 times the number of degrees of freedom (number of flux values used to infer labels) and (iv) to have reasonable uncertainties on velocity and stellar abundances and take only the stars for which median SNR $>$ 10. After these quality cuts, 192 stars remain in our sample (cut down from 217 stars which fall within the colour-magnitude boundaries shown in Figure \ref{fig:cmdall} from a total set of 273 stars with SNR $>$ 10.). These 192 stars are shown in the far right panel of Figure \ref{fig:cmdall}, coloured by metallicity, $\ell_{\mathrm{[Fe/H]}}$. Reassuringly there is a trend in [Fe/H] with colour, with metal-poor stars at left. The scatter in this
gradient is expected given the continuous star formation history  of NGC~6822 (i.e., introducing an expected age-metallicity-colour degeneracy on the red giant branch). We note that of the quality cuts, the $\chi^2$ restriction only eliminates $\approx$10 percent of the stars, and visual inspection reveals that a number of these have significant artefacts in the spectra from imperfect spectra extraction. Therefore, more than 90 percent of the NGC~6822 stellar sample is able to be fit well with the model built on \lamost\ spectra. About 3 percent of the stars that meet our quality cuts fall into slightly un-physical spaces on the red giant branch, with inferred \teff\ labels that fall slightly cool or hot at these \logg\ labels. We do not exclude these few (possibly mislabeled main sequence) stars and they do not impact our overall results.

%final2/makechi2hist.py
\begin{figure}
\centering
\includegraphics[scale=0.28]{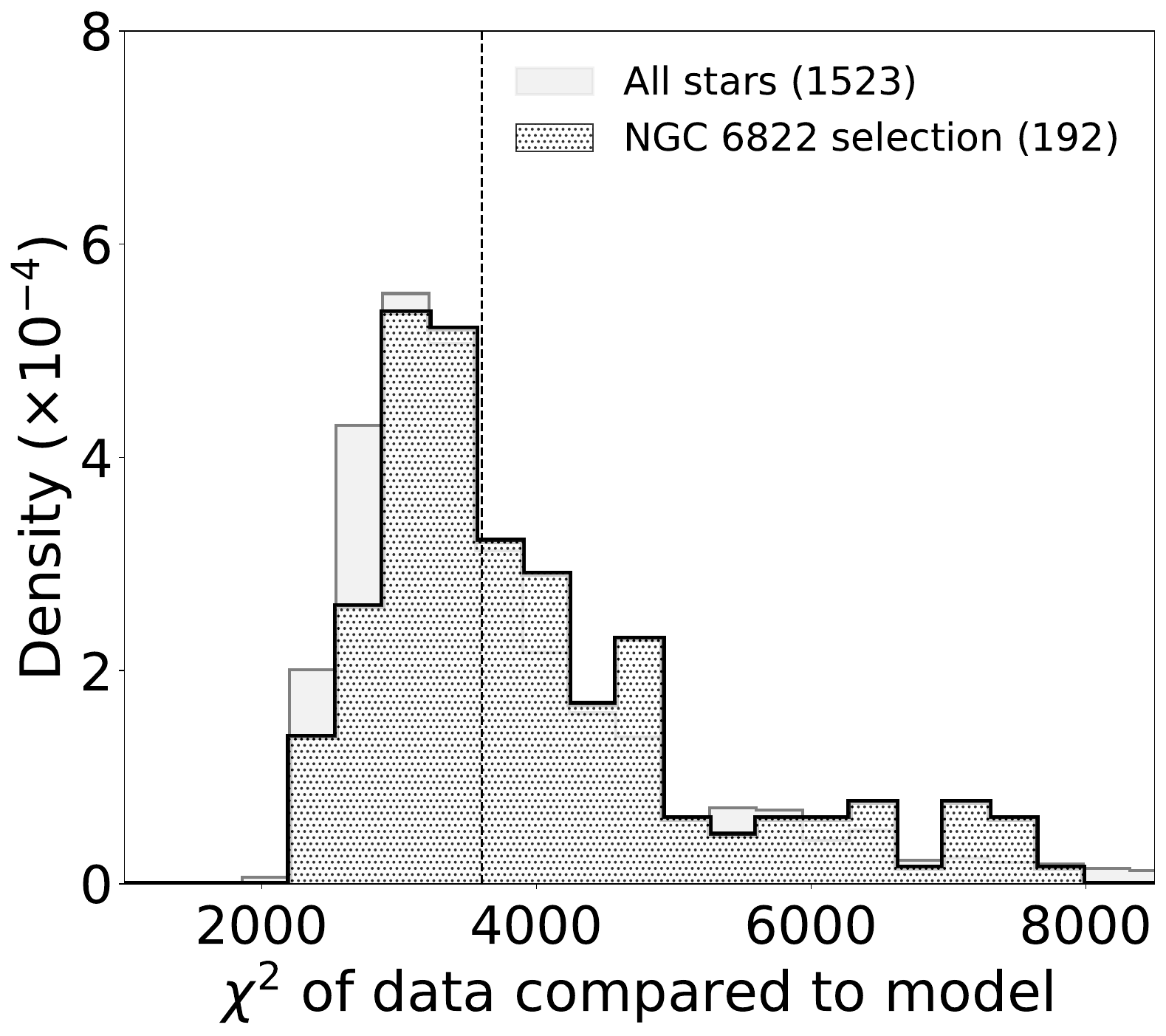}
\caption{The $\chi^2$ distribution for the 1523 resolved IFU spectra extracted from MUSE in the light grey histogram. The dot hatch histogram shows the 192 stars that meet the quality cuts described in the text and Figure five; the median $\chi^2$ of this distribution peaks at 3600, which is slightly larger than the number of degrees of freedom  (wavelength points = 2649) and equivalent to a $\chi^2_{red}$ = 1.35. We note that the flux uncertainties for the MUSE spectra (which go into the denominator of the $\chi^2$ calculation) are likely underestimated.}
\label{fig:chi2}
\end{figure}

The $\chi^2$ distribution of the data compared to the model for each of the spectra is shown in Figure \ref{fig:chi2}. The degrees of freedom is indicated. The grey shaded histogram is for the 192  stars that meet quality cuts described.  The Appendix showcases some individual spectral fits between model and data in Figures \ref{fig:specone}, \ref{fig:spectwo}, \ref{fig:specthree} and \ref{fig:specfour}. The median $\chi^2_{\mathrm{reduced}}$ = 1.35, which likely reflects the (known) underestimate of propagated uncertainties by the MUSE pipeline \citep[][]{Bacon2017}. 

We can obtain a reduced-$\chi^2$ = 1 with flux uncertainties inflated by 17 percent. This subsequently also increases the label errors, by $\sim$30 percent for \teff\ and \logg, $\sim$20 percent for $\ell_{\mathrm{[Fe/H]}}$, $\ell_{\mathrm{[Mg/Fe]}}$ and $\ell_{\mathrm{[Mn/Fe]}}$, $\sim$40 percent for $\ell_{\mathrm{[C/Fe]}}$, $\sim$50 percent for $\ell_{\mathrm{[Al.Fe]}}$, and $\sim$100 percent for $\ell_{\mathrm{[N/Fe]}}$. This corresponds to an inflation of the median uncertainties of the NGC~6822 sample which we report in Table \ref{tab:inflation}.

%final7/makecolormag_multi_four.py  
\begin{table*}
\centering
\footnotesize
\begin{tabular}{|l l l l l l l l l l  |} 
 \hline
inflation factor & $\sigma_{Teff} [K] $ & $\sigma_{logg} [dex] $ & $\sigma_{\mathrm{[Fe/H]}}$ & $\sigma_{\mathrm{[Mg/Fe]}}$ & $\sigma_{\mathrm{[Mn/Fe]}}$ &  $\sigma_{\mathrm{[Al/Fe]}}$ &   $\sigma_{\mathrm{[C/Fe]}}$ &  $\sigma_{\mathrm{[N/Fe]}}$ \\ [0.5ex] 
 \hline\hline
1 & 88 & 0.26 & 0.17 & 0.10 & 0.12 & 0.18 & 0.13 & 0.13\\
\hline
1.17 & 113 & 0.33 & 0.21 & 0.11 &0.15 & 0.27 & 0.18 & 0.29 \\
 \hline\hline
\end{tabular}
\caption{Median uncertainties of the 192 NGC~6822 stars determined from the monte-carlo procedure described in the text given the \lamost\ reported uncertainties (first row) and from the \lamost\  reported uncertainties inflated by 17 percent (second row). }
    \label{tab:inflation}
\end{table*}

\begin{figure}
\centering
\includegraphics[scale=0.28]{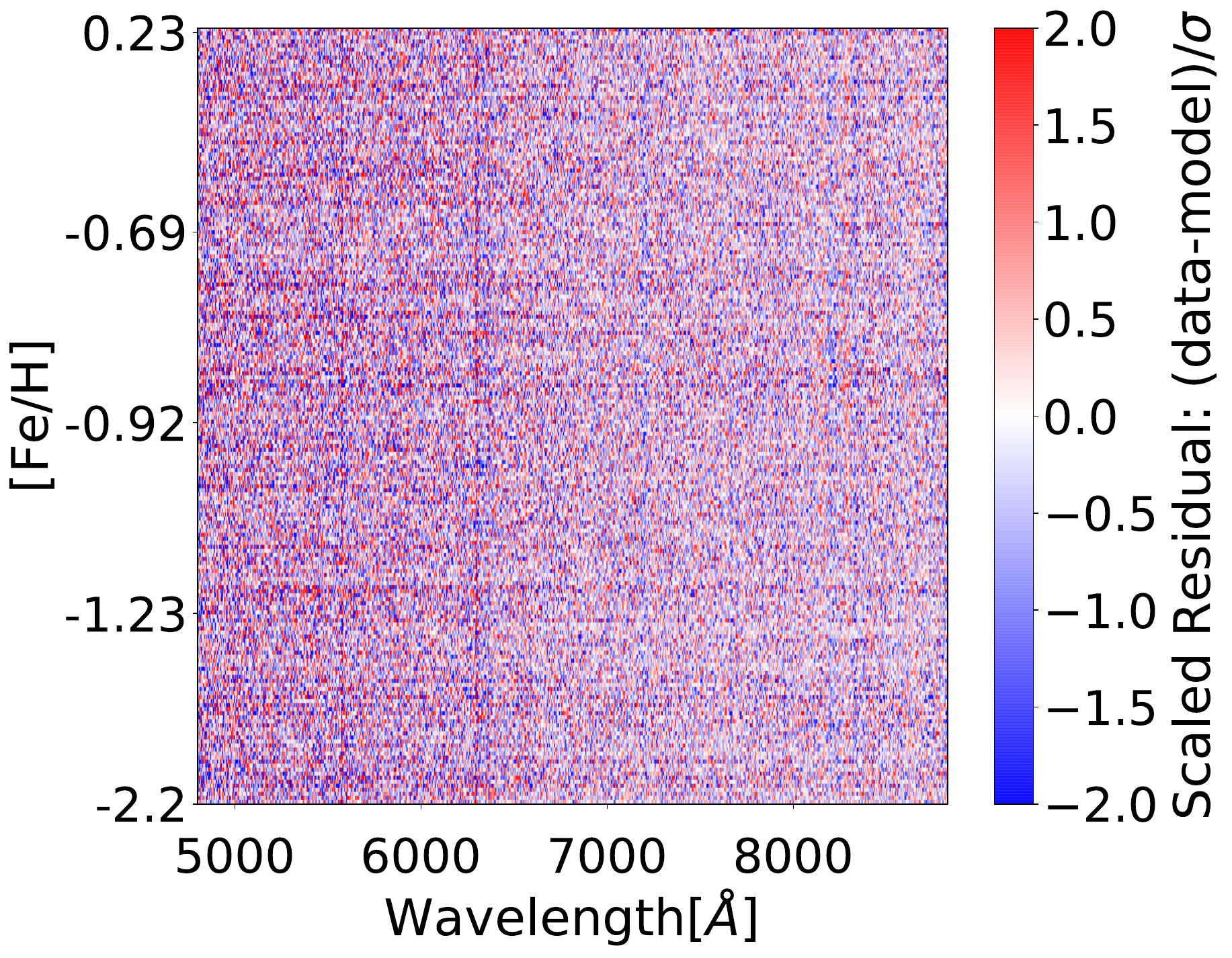}
\caption{A matrix of the  residual amplitude of data-model scaled to the uncertainty (shown in the colourbar) of the 192  stars (shown 1 per row and ordered in [Fe/H]), that meet the quality cuts described in the text. The uncertainty is the quadrature sum of the flux uncertainty and the model's scatter term. The residual mean is zero and the average $1-\sigma$ standard deviation of the (data-model) values alone is $<$ 0.1.}
\label{fig:imshow_matrix}
\end{figure}

A summary of the residuals, ordered by [Fe/H] for all 192  stars is shown in Figure \ref{fig:imshow_matrix}. The residuals are unbiased and have an average $1-\sigma$ standard deviation of $\pm$ 0.10. This shows that the spectrum of NGC~6822 is fit reasonably well (given the median SNR=15, and the $\chi^2$ per flux value indicates the model is a  reasonably good fit to the data). The residuals demonstrate there is not an ensemble of systematic amplitudes showing under- or over-prediction at any features at any wavelength by the model built on Milky Way spectra \citep[for a counter example when there are features not systematically well fit see][]{Rampalli2021}.

%final7/newmodel/revc/makeimshow_scaled2.py

The mean abundance labels we report for NGC~6822 using 192 stars are: $\ell_\mathrm{[Fe/H]} = -0.90 \pm 0.03$, $\ell_\mathrm{[Mg/Fe]} = -0.01 \pm 0.01$, $\ell_\mathrm{[Mn/Fe]} = -0.22 \pm 0.02$, $\ell_\mathrm{[Al/Fe]} = -0.32 \pm 0.03$, $\ell_\mathrm{[C/Fe]} =-0.43 \pm 0.03$, $\ell_\mathrm{[N/Fe]} =0.18 \pm 0.03$. The tabulated values of all labels inferred by applying \tc\ are provided in full online (including uncertainties) and Table \ref{tab:results} shows an excerpt of the label values.

\subsection{Comparison to other studies of NGC~6822}

%final7/newmodel/revb/makeswankirbyhist.py
%final7/newmodel/revc/makeimshow_scaled.py
\begin{figure}
\centering
\includegraphics[scale=0.28]{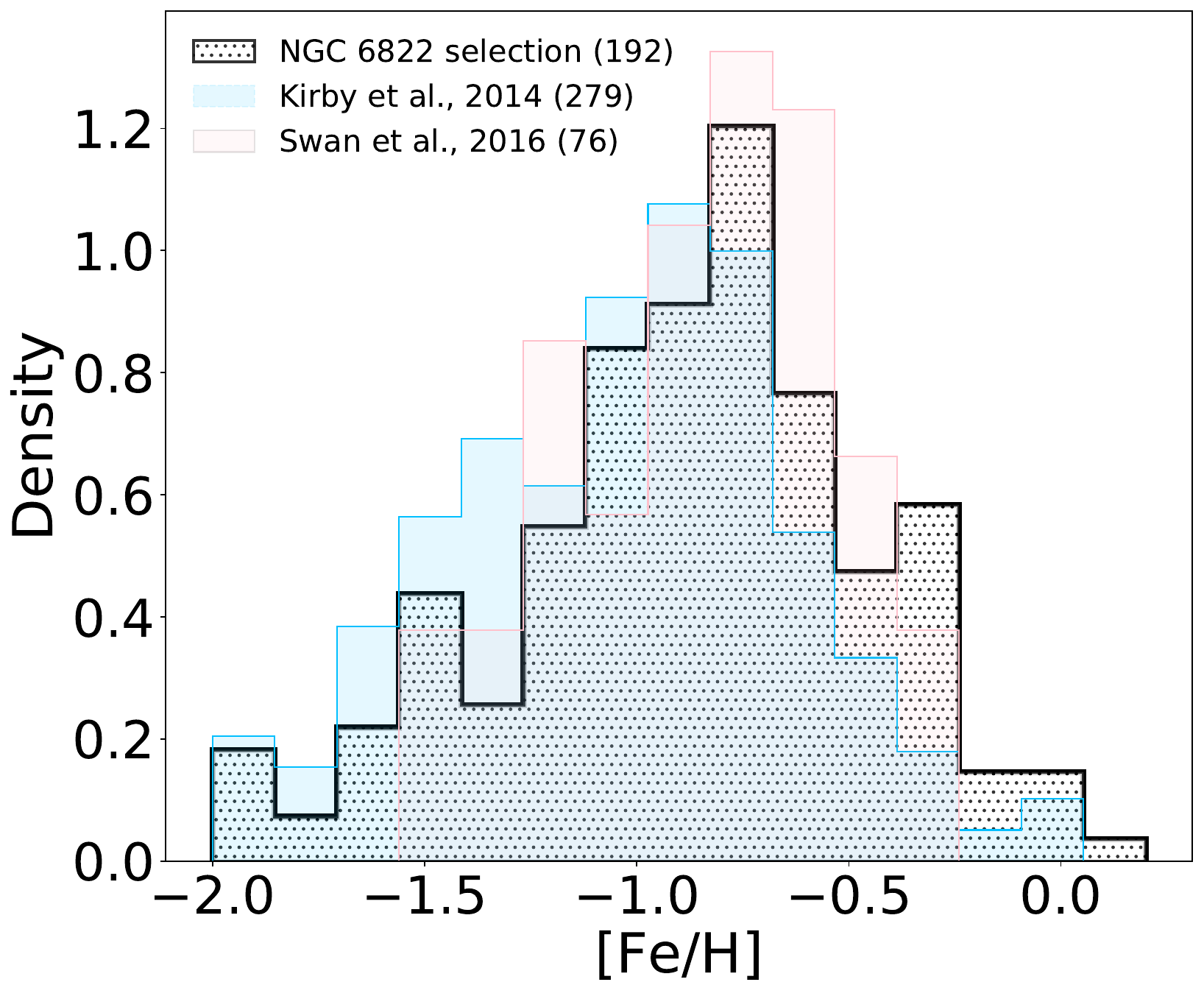}
\caption{A comparison of the inferred [Fe/H] label for our study with that of medium resolution analyses of \citet{Kirby2014} and \citet{Swan2016}, respectively. There is very good agreement between the studies. The number of stars is indicated in brackets.}
\label{fig:fehcompare}
\end{figure}

Although we are careful to state that we undertake a label inference, the element abundances that we find for NGC~6822 are consistent with previous spectroscopic measurements as well as expectations of dwarf galaxies \citep[e.g.][and references therein]{Venn2004, Tolstoy2009, Skul2019}.

We find an $\ell_{\mathrm{[Fe/H]}}$ = --0.90 $\pm$ 0.03, which compares well with spectroscopic studies of similar stellar populations. (Note the uncertainty quoted here is statistical confidence on the mean whereas the standard deviation of the population of 192 stars in NGC~6822 is 0.46 dex). The spectroscopic analysis of medium resolution data by \citet{Kirby2013} determines a mean [Fe/H] = $-1.1$ $\pm$ 0.03 (statistical uncertainty, with a standard deviation of $\pm$0.5) from observations of 279 red giant stars in NGC~6822. The stars in the \citet{Kirby2013} sample are distributed across the entire galaxy and the three in the field of view of the MUSE footprint that are not in common with our study. There is also a reported gradient that implies the mean metallicity is spatially dependent of $-0.28 \pm 0.08$~dex/kpc \citep{Taibi2022}.
The medium-resolution analysis of 76 stars from \citep{Swan2016} finds a mean [Fe/H] = $-0.84$ $\pm$ 0.02. The comparison of the three samples is shown in Figure \ref{fig:fehcompare}.

  We do obtain a lower metallicity for the NGC~6822 disk to that measured from high resolution 8-m class telescope data of super-giant stars. \citet{Venn2001} find an [Fe/H] = $-0.49 \pm 0.22$ and \citet{Patrick2015} find a mean Z=$-0.52$, using super-giants as targets. This is most likely because the stellar populations are young for the most luminous giants targeted in these studies and have likely experienced a more advanced chemical enrichment than the relatively older stars.  We also note that the NGC~6822 photometric metallicity estimates are [Fe/H] $\sim$ -1.0 \citep{Fusco2014, Dav2003}.

\subsection{Comparison to the Milky Way}

%final7
%run -i sampleit_one.py does not sample
\begin{figure*}
\centering
\includegraphics[scale=0.55]{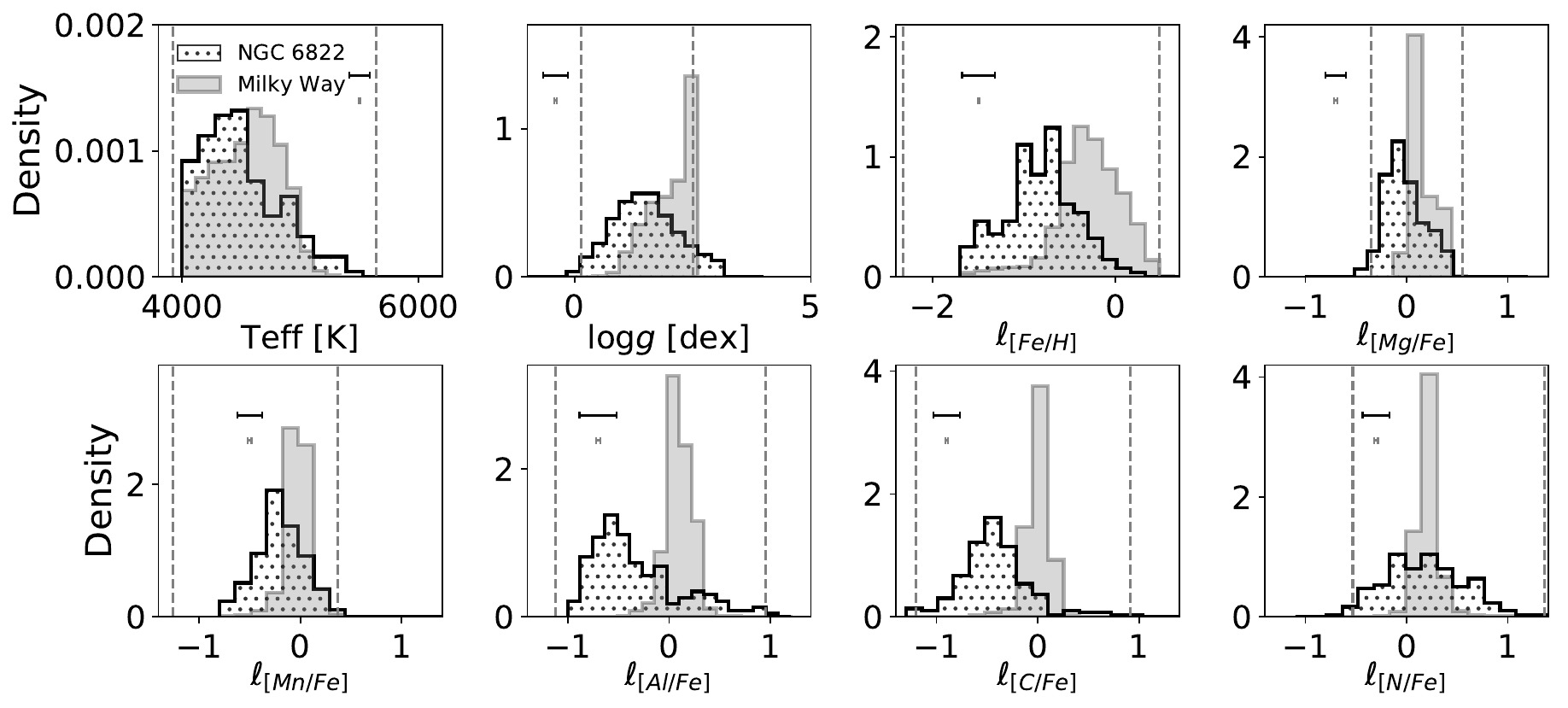}
\caption{A comparison of the inferred [Fe/H] label for our study with that of the Milky Way for 182,800 Milky Way stars with the quality cuts used to make the reference objects. NGC~6822 is notably lower than the Milky Way in all element labels apart from $\ell_{\mathrm{[N/Fe]}}$. Median error bars are shown for the two samples; the data for the Milky Way has a factor of 10 higher SNR than NGC~6822 (at 175 compared to 15) and the resolution is a factor of 12 higher (R=22,500 compared to 1800). The uncertainties are a factor of $\sim$7-10 higher for NGC~6822, with the exception of $\ell_{\mathrm{[Fe/H]}}$ which is $\sim$18 times lower for \apogee. The dashed vertical lines shown the boundaries of the reference objects used for training the model with \tc.}
\label{fig:mwcompare}
\end{figure*}

The labels for NGC~6822 are on the scale of the \apogee\ survey and we therefore can directly contrast the abundances for the Milky Way and NGC~6822. In Figure \ref{fig:mwcompare} we show the abundance distributions of NGC~6822 compared with those of the \apogee\ survey for 182,000 red giant stars in \apogee\ with shared  properties as the training objects. That is, red giant stars with \logg\ $<$ 2.5 and $\mathrm{[X/Fe]}_{err}$ $<$ 0.2, as well as an additional criterion of a distance from the Sun of $<$ 10~kpc. The distances are taken from the value added ASTRONN catalogue \citep{Leung2019}.  We highlight that the measurements are able to be compared and contrasted, even though the data quality between \apogee\ and \lamost\ is vastly different. The \apogee\ labels are derived from R=22,500 spectra and the mean SNR of the stars shown in Figure \ref{fig:mwcompare} is  SNR = 245. The NGC~6822 labels are derived from the \lamost\ R=1800 spectra with a median SNR = 15. The mean label measurements are able to be directly compared; each has a high precision of $(\sigma_{\ell_{\mathrm{[X/Fe]}}}/\sqrt N$) where N is the number of stars. However, as the measurement uncertainties are different between the two samples, the spreads seen for each element abundance are not directly comparable between the two galaxies. The reported \apogee\ uncertainties are a factor of 10 or more lower than those for NGC~6822. The width of the distributions are near-intrinsic for the Milky Way, whereas for NGC~6822 they are broadened by the larger uncertainties. The mean uncertainties on the labels are indicated in each sub-figure, with the NGC~6822 error-bars above those of \apogee.  Typical label uncertainties of the NGC~6822 stars are: $\sigma_{err}{_{(\ell_{\mathrm{[Fe/H]}})}} = 0.17$,
$\sigma_{err}{_{(\ell_{\mathrm{[Mg/Fe]}})}} = 0.10$,
$\sigma_{err}{_{(\ell_{\mathrm{[Mn/Fe]}})}} = 0.12$,
$\sigma_{err}{_{(\ell_{\mathrm{[Al/Fe]}})}} = 0.18$,
$\sigma_{err}{_{(\ell_{\mathrm{[C/Fe]}})}} = 0.13$,
$\sigma_{err}{_{(\ell_{\mathrm{[N/Fe]}})}} = 0.13$.
Similarly to other dwarf galaxies, the mean abundances for these elements is lower than that of the mean abundance of the Milky Way \citep[see][and references therein]{Venn2004, Tolstoy2009, Ji2023, Skul2019}.

\begin{figure*}
\centering
\includegraphics[scale=0.5]{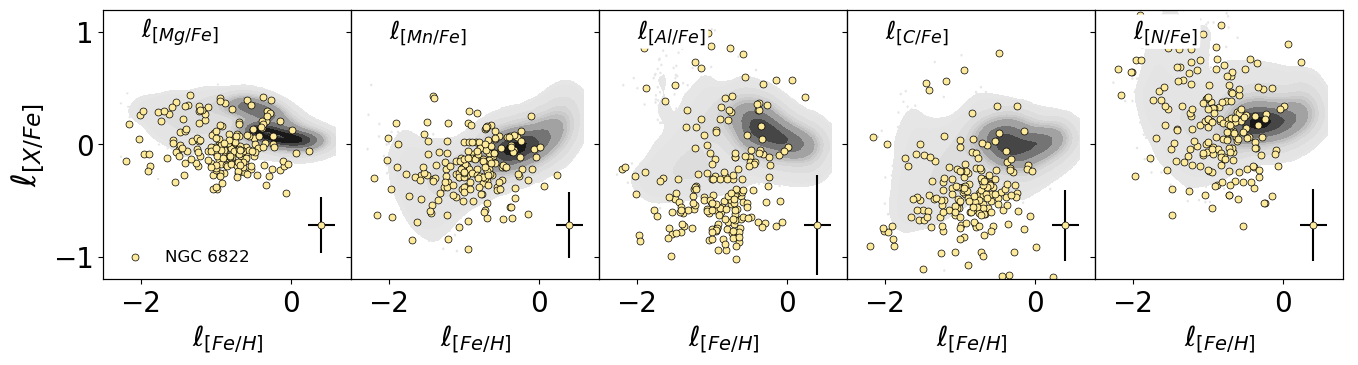}
\caption{The Milky Way element abundance distribution shown in greyscale, and the 192 NGC~6822 stars, in yellow.}
\label{fig:ecc1}
\end{figure*}

Dwarf galaxies are believed to be the building blocks of the Milky Way at low-metallicity \citep[e.g.][]{Searle1978} and new element abundance and kinematic data has accelerated a reconstruction of the assembly history and progenitors \citep{Mackereth2019, Naidu2020, Helmi2020, Belokurov2023, Horta2023, Han2024, Horta2024,  Sestito2024a, Sestito2024b}.  Dwarf galaxies show distinct trends compared to the Milky Way. However, the local dwarf galaxy element abundances and the low-metallicity stars of the Milky Way at [Fe/H] $<$ -1.0 clearly show some overlap \citep[e.g. see][for a direct comparison]{Hasselquist2021}.  High-mass dwarf galaxies like the Large Magellanic Cloud (LMC) and NGC~6822 also show an extended distribution to metallicities [Fe/H] $\sim$ -0.5. If analogue  systems are contributors to the Milky Way at low-metallicity, their metal-rich material presumably also may reside in the abundance space of the Milky Way disk. In Figure \ref{fig:ecc1} we show the 192 NGC~6822 stars (yellow circles) as well as the Milky Way population reported in Figure \ref{fig:mwcompare} in greyscale. The typical error bars on the NGC~6822 measurements are included in each sub-figure.

%makeecc_xy3_pt7.py 
%makeecc_xy3_pt7_all.py 
%run -i test6.py

%makeabundplots_dwarfs_density.py
%run -i test8
%run -i test8_mean
\begin{table*}
\centering
\footnotesize
\begin{tabular}{| c c c c c c c c c |} 
 \hline
Dwarf Galaxy & SNR & [Fe/H] & [Mg/Fe] & [Mn/Fe] & [Al/Fe] & [C/Fe] & [N/Fe] & No. stars   \\
 \hline
 \hline
NGC~6822 & 15 & -0.9 $\pm$0.03 & -0.01  $\pm$0.01  & -0.22  $\pm$0.02  & -0.33  $\pm$0.03  & -0.43  $\pm$0.03  & 0.18  $\pm$0.03 & 192\\ 
 \hline
LMC & 91 & -0.83 $\pm$0.01 & 0.03 $\pm$0.01 & -0.18 $\pm$0.01 & -0.32 $\pm$0.01 & -0.41 $\pm$0.01 & 0.2 $\pm$0.01 & 2522 \\
 \hline
SMC & 87 & -1.04 $\pm$0.01 & -0.01 $\pm$0.01 & -0.24 $\pm$0.01 & -0.43 $\pm$0.01 & -0.47 $\pm$0.01 & 0.2 $\pm$0.01 & 1284 \\
 \hline
Sculptor& 64 & -1.49$\pm$0.03 & -0.07$\pm$0.02  & -0.21$\pm$0.03 & -0.68$\pm$0.02&  -0.73$\pm$0.03 & 0.31$\pm$0.04& 103 \\
 \hline
Fornax & 56 & -1.01$\pm$0.02 &  -0.15$\pm$0.02 &  -0.31$\pm$0.02 &  -0.61$\pm$0.02 & -0.58$\pm$0.02 & 0.17$\pm$0.02 & 174 \\ [1ex] 
 \hline
 \end{tabular}
 \caption{A summary of the mean abundances for four \apogee\ dwarf galaxy abundances as well as NGC~6822 as measured in this study, and shown in Figure \ref{fig:apcompare}. The uncertainties reported are the confidence on the mean values.}
  \label{tab:tabletwo}
%\end{center}
\end{table*}

\subsection{Comparison of NGC~6822 to other Dwarf Galaxies}

Dwarf galaxies individually show distinct tracks in element abundances,  [X/Fe], but have shared mean properties compared to the Milky Way, and span a range of [Fe/H] \citep[e.g.][]{ Monaco2005, McWilliam2006, Sbordone2007, Letarte2010, Kirby2010, Venn2012, Norris2017a, Norris2017b,  Lemasle2014,Shetrone2003,Hendricks2014, Skul2019, Skul2020, Reichert2020, Hasselquist2021, Tang2023, Tolstoy2009, Kirby2013, Escala2020, Ji2023, Frebel2023}.

%makeabundplots_dwarfs_density.py
%run -i test8
\begin{figure*}
\centering
\includegraphics[scale=0.55]{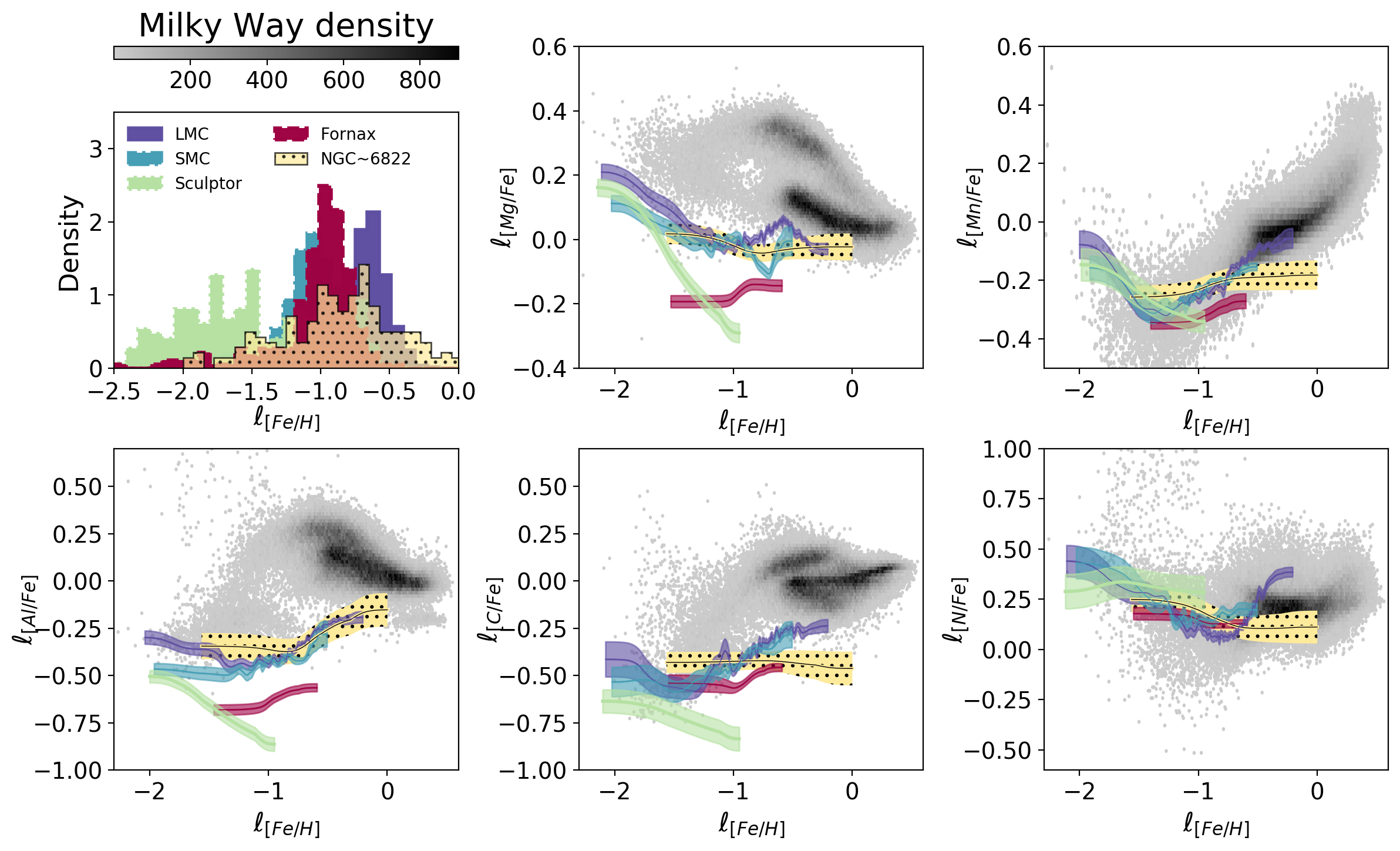}
\caption{The running median  of the element abundance labels [X/Fe] for four dwarf spheriodal galaxies observed by \apogee\ as well as NGC~6822 as reported here, as well as the Milky Way stellar distribution. The Milky Way is shown in greyscale. The shaded regions show the confidence on the mean value. The left top sub-figure shows the comparison of the [Fe/H] of the dwarf spheriodals under consideration, with NGC~6822 shown in hatched yellow shading.  This comparison shows that the measurements for NGC~6822 are consistent with other dwarf galaxies for all element abundance labels.}
\label{fig:apcompare}
\end{figure*}

The \apogee\ survey has targeted a set of massive local group galaxies which are analysed in detail in \citet{Hasselquist2021}. We can directly compare these dwarf galaxy abundances measured by \apogee\ to the measurements we make of NGC~6822, as they are on the same label scale.  We show the Milky Way stellar distribution along with the median [Fe/H]-[X/Fe] trends for four \apogee\ dwarf galaxies and NGC~6822 in Figure \ref{fig:apcompare}. We do this to demonstrate that our inferred labels sit in the parameter space of dwarf galaxies. We examine Fornax, the Large Magellanic Cloud (LMC), the Small Magellanic Cloud (SMC) and Sculptor. We select the stars in these galaxies using the \texttt{FIELD} column from the \apogee\ data release and adopt the velocity cuts in \citet{Hasselquist2021}. In addition we only consider stars with logg $<$ 2.5. 

The summary of mean abundance labels for the \apogee\ dwarfs and NGC~6822 is given in Table \ref{tab:tabletwo}. This analysis demonstrates that all of our abundance labels follow expectations of dwarf galaxies and despite having substantially lower quality data we obtain abundance labels of good quality with comparable labels and precision on the mean abundances.

The dwarf galaxy IC 1613 is a close analog to NGC~6822 with a similar mass and star formation history, and isolated. Using 275 red giant branch stars, resolved from MUSE IFU data, \citet{Taibi2024} report a mean [Fe/H] of -1.06 (with an intrinsic 1-$\sigma$ standard deviation of 0.26 dex). This is obtained using the Calcium triplet metallicity calibration \citep{Starkenburg2010} and is again similar to the metallicity $\ell_{\mathrm{[Fe/H]}}$ that we infer for NGC~6822. 

\section{Discussion} 

We have demonstrated nominal success in labelling NGC~6822 spectra with stellar parameters and abundances, using a model built on the \apogee\ survey labels and \lamost\ spectra. This approach shows that the spectral data can be reasonably well fit with the inferred labels, despite the issue that the observations are performed on different instruments and with different data reductions. Other studies have also shown good performance when this is the case \citep[e.g.][]{Rice2020, Wang2022}. There are residual telluric and skyline artefacts in the data as shown in the individual spectra in the Appendix (see Figures \ref{fig:specone}, \ref{fig:spectwo}, \ref{fig:specthree}, and \ref{fig:specfour}). These do not have obvious impact on the label precision, presumably as the model form is very constrained. The labels can not vary in a way to fit these clearly anomalous and substantial features such at the model fit across wavelength improves. Our results for NGC~6822 show good agreement with other spectroscopic studies of red giant stars as shown in Figures \ref{fig:fehcompare} and \ref{fig:ecc1}. Furthermore, the abundance labels are comparable to those of dwarf galaxies measured by \apogee\ using high resolution higher SNR data. The top of red giant branch, where are our stars extend to is hard to model ab initio. We note that this means we should be cautious about the accuracy of the training labels (and the labels we subsequently infer), but also motivates using a data-driven model to take advantage of the complex information in the spectrum.

\subsection{Label Validation, Tests and Assumptions over Wavelength}

We test the label inference over a number of different wavelength regions. We find small variations in mean abundance labels using restricted wavelength regions, of up to $\sim$0.1~dex, and the uncertainties increase as the number of wavelengths used decreases. Using different regions however, the overall comparative results and conclusions remain unchanged.  We undertake two tests to report comparisons of to demonstrate the model performance. First, excluding wavelengths $<$ 6000\AA, and second, excluding wavelengths $>$ 6000\AA. For the first test, whereby the labels are learned only from $>$ 6000\AA, we find that the inferred abundance labels are as follows: 
$\ell_\mathrm{[Fe/H]} = -0.80 \pm 0.05$, $\ell_\mathrm{[Mg/Fe]} = 0.01 \pm 0.02$, $\ell_\mathrm{[Mn/Fe]} = -0.22 \pm 0.03$, $\ell_\mathrm{[Al/Fe]} = -0.19 \pm 0.04$, $\ell_\mathrm{[C/Fe]} =-0.45 \pm 0.04$, $\ell_\mathrm{[N/Fe]} =0.26 \pm 0.04$. For the second test, whereby the labels are learned only from $<$ 6000\AA, we find that the inferred labels are as follows: $\ell_\mathrm{[Fe/H]} = -0.87 \pm 0.05$, $\ell_\mathrm{[Mg/Fe]} = 0.02 \pm 0.02$, $\ell_\mathrm{[Mn/Fe]} = -0.18 \pm 0.03$, $\ell_\mathrm{[Al/Fe]} = -0.19 \pm 0.04$, $\ell_\mathrm{[C/Fe]} =-0.33 \pm 0.04$, $\ell_\mathrm{[N/Fe]} =0.20 \pm 0.04$. Uncertainties become prohibitively large when using less than 20 percent of the spectral region for label inference. 

 As done in numerous prior studies,  we allow \tc\ to use the information across the entire spectral region. In some studies we have tested censoring around individual element lines only, to validate the same results are obtained (but with subsequently higher uncertainties) and to demonstrate the model fits the data at specific atomic features being labeled \citep[i.e.][]{Manea2023}. This test is not similarly feasible at low-resolution for all the elements as these are blended.  However, we do test and validate that the spectrum is a good fit to the data, including at individual strong line features including the Ca-triplet and Mg-lines (see  the Appendix, Figures \ref{fig:model2} \ref{fig:specone}, \ref{fig:spectwo}, \ref{fig:specthree}).

\section{Conclusion}

We have demonstrated that we can employ \tc\ with a model trained on \lamost\ spectra and labels to infer stellar parameters and abundances for extra-galactic spectra at R=1800 and SNR $>$ 10. As seen in prior studies, \tc\ works well for low-resolution data where individual element features are typically blended and resolved only for a few strong lines. However, in most previous implementations of \tc, the reference set of stars used to build the model and the test set of stars to be labelled are observed with the same instrument. This is not satisfied in this study. Nevertheless, the labels that are inferred generate a model that is a reasonably good fit to the NGC~6822 spectra and the abundance labels show amplitudes consistent with prior studies and the expectations of dwarf galaxies. For NGC~6822, we report an overall mean $\ell_{\mathrm{[Fe/H]}}$ = -0.9 $\pm$ 0.03 and individual element abundances that qualitatively align with other dwarf galaxies.

Our validation tests give us confidence that our results for NGC~6822 are sensible.  We propose that using \tc\ in this context provides an opportunity to make use of the available IFU data for extra-galactic archaeology using resolved stars. We demonstrate this method using MUSE, but this is transferable to other IFU data including from Local Volume Mapper \citep{Kollmeier2017,LVM} and future instruments like MAVIS \citep{MAVIS,MAVIS2}.  

We caution that although the model we use can find a good fit to the NGC~6822 spectra, this approach may fail to well fit other galaxy populations or label populations that have true abundance amplitudes far outside the ranges of the reference objects used to train the model. We also caution that we do not have isolated atomic features to validate the model fits isolated features of each element, as can be done at higher resolution  \citep{Manea2023}. Therefore, to assess the accuracy of our abundance labels and support our ACACIAS program, complementary high resolution Extremely Large Telescope observations of small numbers of stars in local systems will be an important complement \citep[e.g.][and references therein]{Tolstoy2009, Hasselquist2021}. 

\section*{Acknowledgements}
    Based on data obtained from the ESO Science Archive Facility with DOI: \url{https://doi.eso.org/10.18727/archive/42}.

SB acknowledges support from the Australian Research Council under grant number DE240100150.

MKN thanks Thomas Nordlander  for useful discussions.

KAV thanks the Natural Sciences
and Engineering Research Council of Canada for support through their Discovery and CREATE programs

A.P.J. acknowledges support from the Distinguished Visitor Program by the Research School of Astronomy and Astrophysics at the Australian National University and support from NSF AST-2206264.

 ES acknowledges funding through VIDI grant "Pushing Galactic Archaeology to its limits" (with project number VI.Vidi.193.093) which is funded by the Dutch Research Council (NWO). This research has been partially funded from a Spinoza award by NWO (SPI 78-411).
 
 \section*{Data Availability}
 
 MUSE data for NGC 6822 were taken as part of science verification with program ID 60.A-9348(A), and both raw and processed data products are available through the ESO Science Archive Facility. The resolved MUSE stellar spectra will be shared on reasonable request to the corresponding author.  
 A table of the derived data products (stellar parameters and abundance labels for 192 stars in NGC 6822) is made available in an online table in this work.

\appendix

\section{Element label uncertainties}

\begin{figure}
\centering
\includegraphics[scale=0.35]{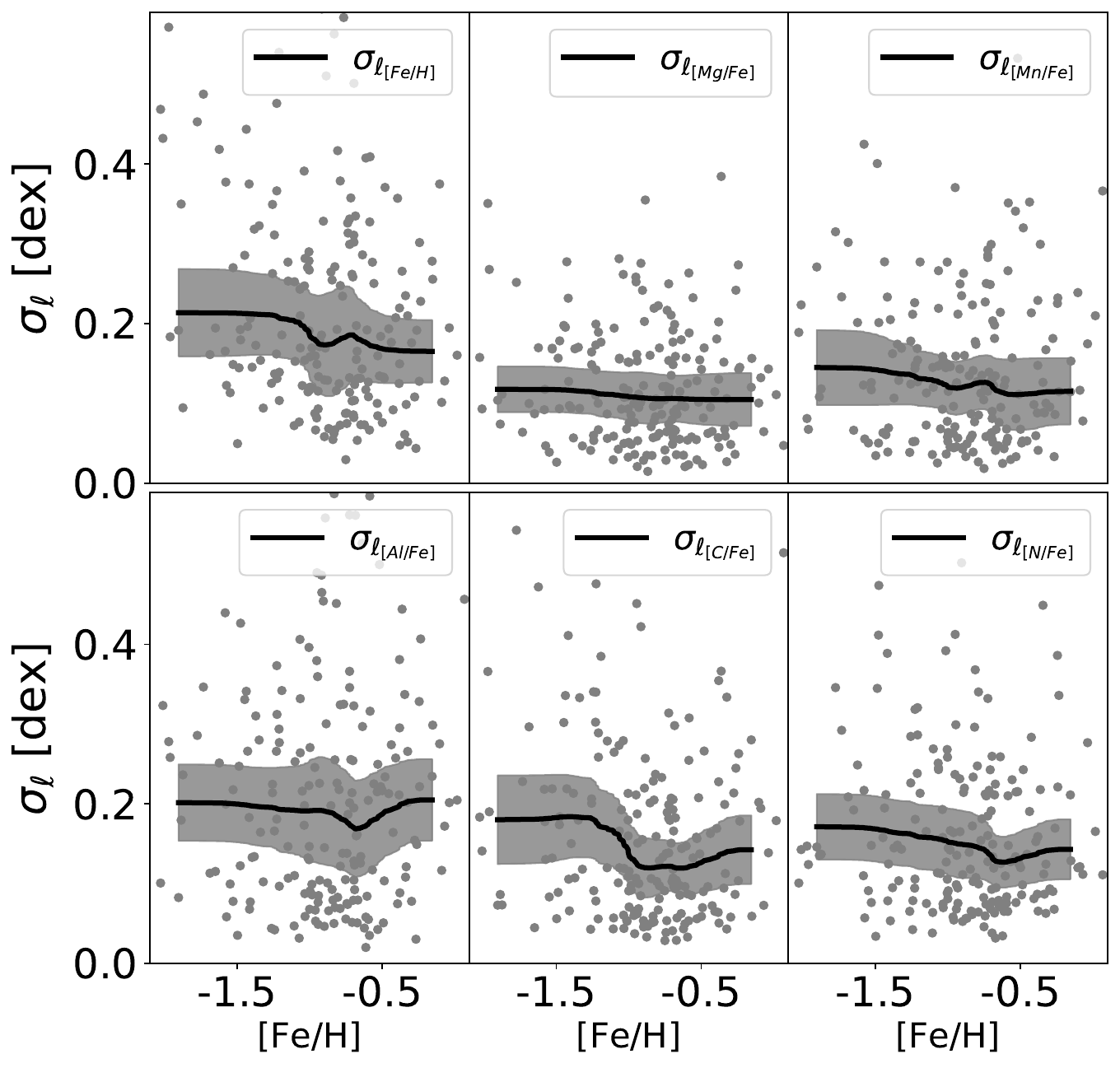}
\caption{The uncertainty of the inferred abundances for the 192 NGC~6822 stars that meet our analysis criteria as a function of [Fe/H]. Individual points are shown to highlight that there are a large range of uncertainties]. }
\label{fig:precision2}
\end{figure}

In Figure \ref{fig:precision2} we show the individual abundance label uncertainties and running median of these uncertainties as a function of [Fe/H]. The mild increase we see in abundance label uncertainty with decreasing $\ell_{\mathrm{[Fe/H]}}$ is consistent with \citet{Leung2019}. The majority of our inferred label values are $\ell_{\mathrm{[Fe/H]}}$ $>$ -2.0. However, below this value the uncertainties on the abundance labels are expected to rapidly increase \citep{Leung2019}.

In contrast to typical abundance measurements using individual line absorption features, most of the individual element enhancements have uncertainties that are lower than that of [Fe/H]\footnote{Fe is used as an overall metallicity indicator in spectroscopy due to the presence of many Fe spectral features}.  This behaviour reflects the full spectrum fitting nature of this approach, as emphasised with our nomenclature. Figure \ref{fig:model} shows that the labels with the highest uncertainties, $\ell_{\mathrm{[Fe/H]}}$ and $\ell_{\mathrm{[Al/Fe]}}$ show high gradient spectra amplitudes distributed across many features, and also particularly at low wavelengths, where the noise is the highest. In contrast, the gradient spectra from other element labels shows lower amplitudes, and lesser information being obtained from the lowest wavelengths.

%test6_cmap.py
%run -i runmeanerr_all_feh2.py  

\section{Alignment between Model and Element Features}

With \tc\ we infer labels from the spectra using all flux amplitudes, and do not restrict the model to any subset of the wavelength values. Element information can manifest in a spectrum beyond the individual absorption features \citep[e.g.][]{Phillips2024}; it would be subsequently interesting to examine how the model coefficients correlate with theoretical spectra synthesised from a model grid of varying abundances, when these become available. Here we examine the first-order origin of the element information in the spectra, namely the element absorption features alone, and test if and how they align with the model's highest amplitude coefficients. The amplitude of the first-order coefficients can serve as a ranking for where the information pertaining to the label is present the spectra. 

\begin{figure*}
\centering
\includegraphics[scale=0.5]{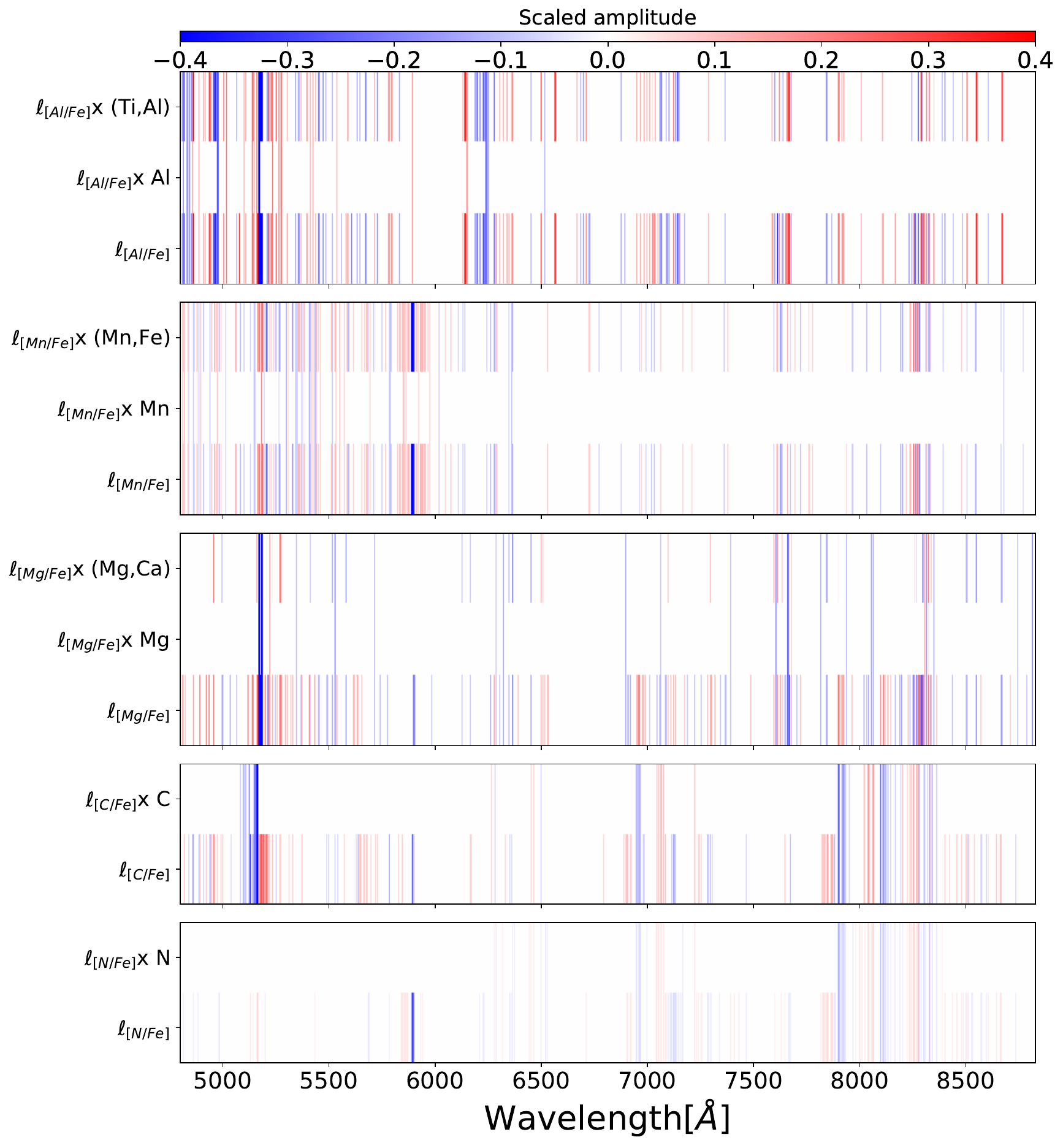}
\caption{A summary of a set of the highest 20 percent of the linear coefficients of \tc's model (bottom row of each of the three sub-figures). The second and third rows for the sub-figures show the projected coefficients from the bottom rows, where element features are present for the elements listed on the y-axis.  This demonstrates that model is using information where there are the relevant element lines (blended at this resolution and not able to be differentiated individually in the spectra itself). Wavelengths are in vacuum. For further discussion of this see the Appendix.}
\label{fig:model2}
\end{figure*}

Figure \ref{fig:model2} examines the relationship between the highest 20 percent amplitude first-order coefficient values and relevant element absorption features.  The bottom row of each sub-figure in Figure \ref{fig:model2}  reports the coefficient amplitudes from Figure \ref{fig:model} for Al, Mn, Mg, C and N, from top to bottom. Note the coefficients are far sparser than shown in Figure \ref{fig:model} simply because only the 20 percent highest absolute values are shown. The element linelist used to identify the absorption features has been downloaded from the VALD atomic database\footnote{$http://vald.astro.uu.se/$} using the stellar extract feature for a reference star of \teff = 4000K, \logg = 1.5 dex, [M/H] = $-0.5$ for line detection threshold $>$ 5 percent. 
The second row of each sub-figure shows the coefficient masked apart from the center wavelengths of the element inferred (in vacuum wavelengths) $\pm$  a resolution element, calculated as $(R/\lambda)$, where R=1800. The third row, for sub-figures with three rows, is similar to the second, but includes an additional element. The second and third panels test if the coefficient amplitude present at that wavelength aligns with the absorption features being labeled or from the same element family as that being labeled. This demonstrates that both element absorption features as well as correlations with are also leveraged. The sub-panel third from bottom shows that the first order [Mg/Fe] label coefficient is aligned with with the Mg features at the Mg triplet ($\sim$5168-5185), and the first-order coefficients for the [Mg/Fe] label shown alignment with other $\alpha$-element lines of e.g. Ca. The sub-figure panels for Mn and Al similarly show the coefficients correspond to the element absorption features for both the element being labeled as well as  elements produced via the same mechanism (e.g. Fe and Ti, respectively). The bottom two sub-figures show the coefficients for the labels [N/Fe] and [C/Fe] align in many cases with absorption features of these elements, which are often together in molecules. There is lesser alignment for N compared to C. There is an expected evolutionary trend with \logg\ up the red giant branch that we see for the labels of [C/Fe] and [N/Fe], so those abundances might be tied to that intrinsic correlation; high gradient spectra (coefficients) for the labels of N and C overlap with logg-sensitive lines (e.g. the Na doublet), see Figure \ref{fig:model}.

Note that negative (blue) coefficients reflect where an increasing label amplitude is correlated with an increasing absorption depth. If \tc\ learns an abundance directly from that element absorption feature itself the expectation is that the coefficient is negative.

%run -i makecmap.py
%run -i makemodelspectrum2_readin.py and makemodelspectrum2_readin_two.py
%makemodelspectrum2_readin.py
%run -i makemodelspectrum2_movecbar2.py   

\section{Examples of model versus data}

\begin{figure*}
\centering
\includegraphics[scale=0.36]{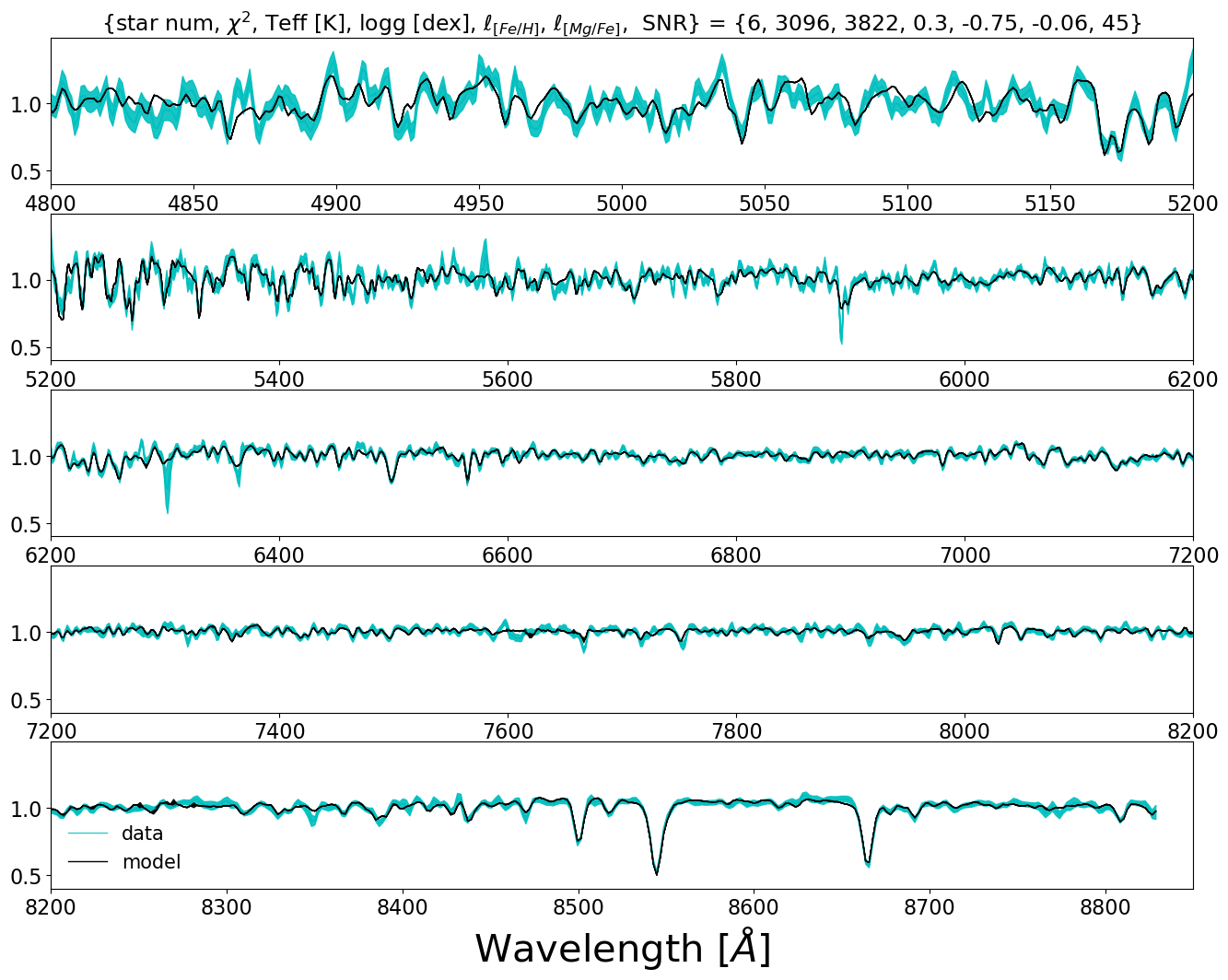}
\includegraphics[scale=0.36]{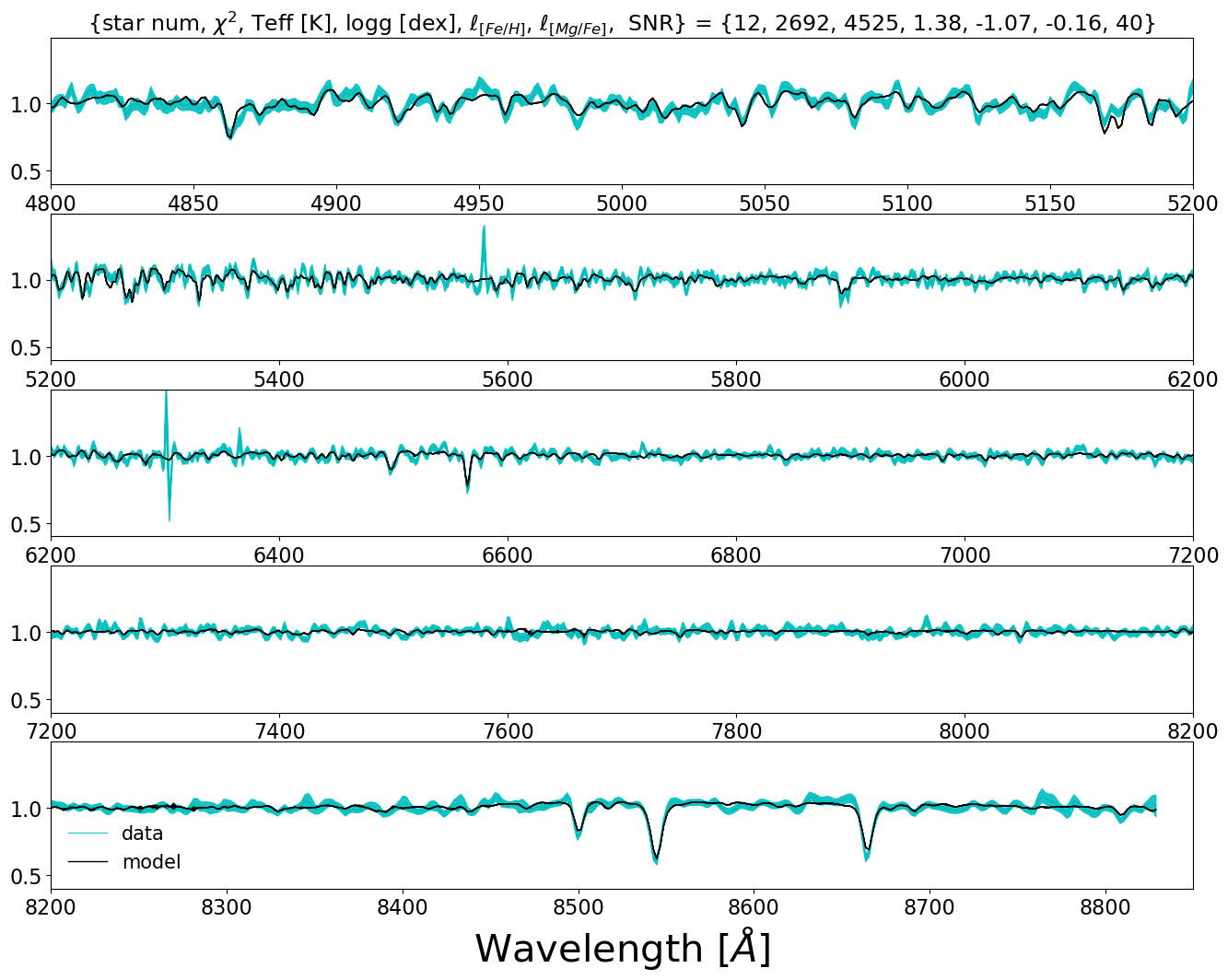}
\caption{Two example fits of the model (black) and data (cyan) for stars with 30 $<$ SNR  $<$ 50. Summary labels are indicated at top.}
\label{fig:specone}
\end{figure*}

\begin{figure*}
\centering
\includegraphics[scale=0.36]{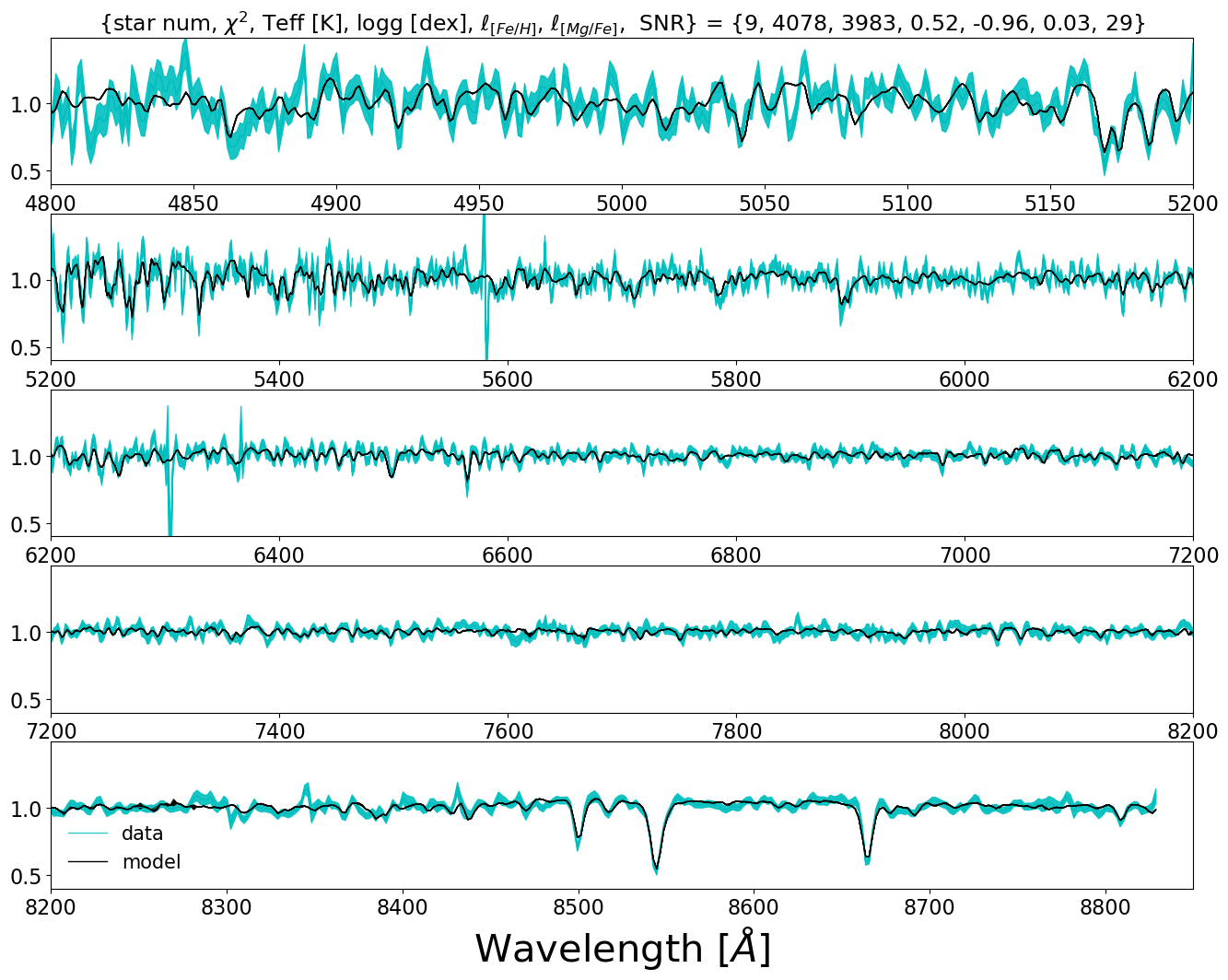}
\includegraphics[scale=0.36]{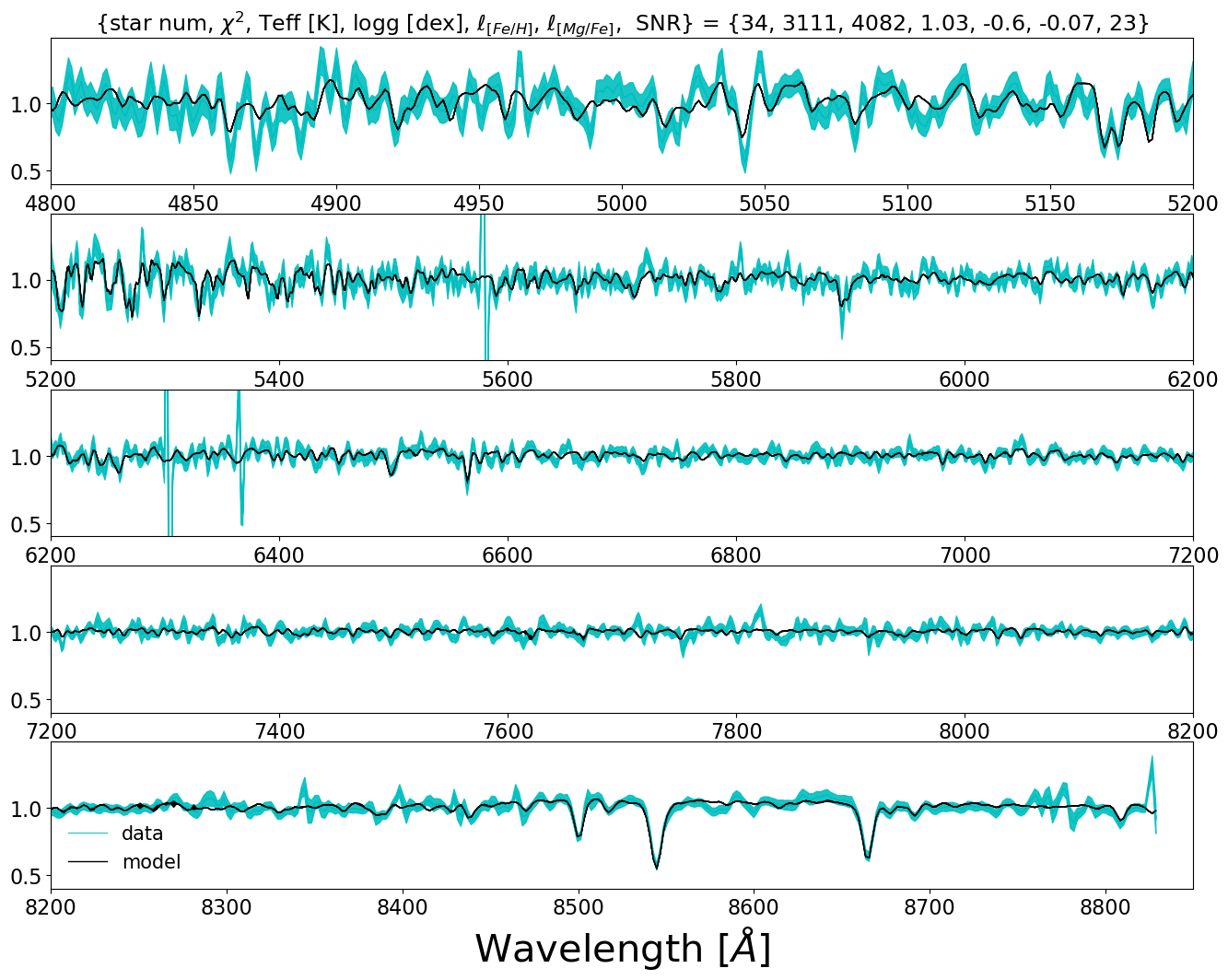}
\caption{Two example fits of the model (black) and data (cyan) for stars with 20 $<$ SNR  $<$ 30. Summary labels are indicated at top.}
\label{fig:spectwo}
\end{figure*}

\begin{figure*}
\centering
\includegraphics[scale=0.36]{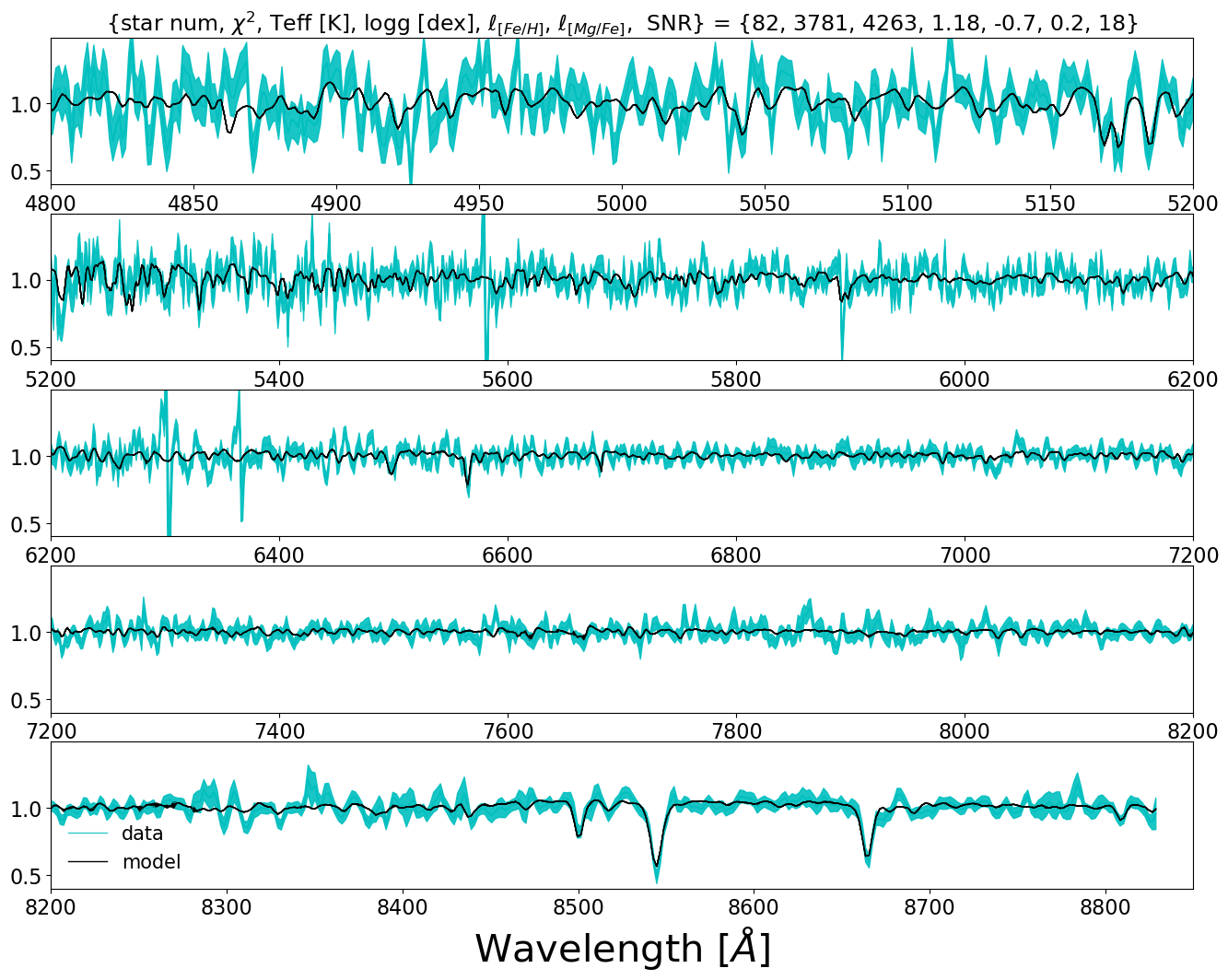}
\includegraphics[scale=0.36]{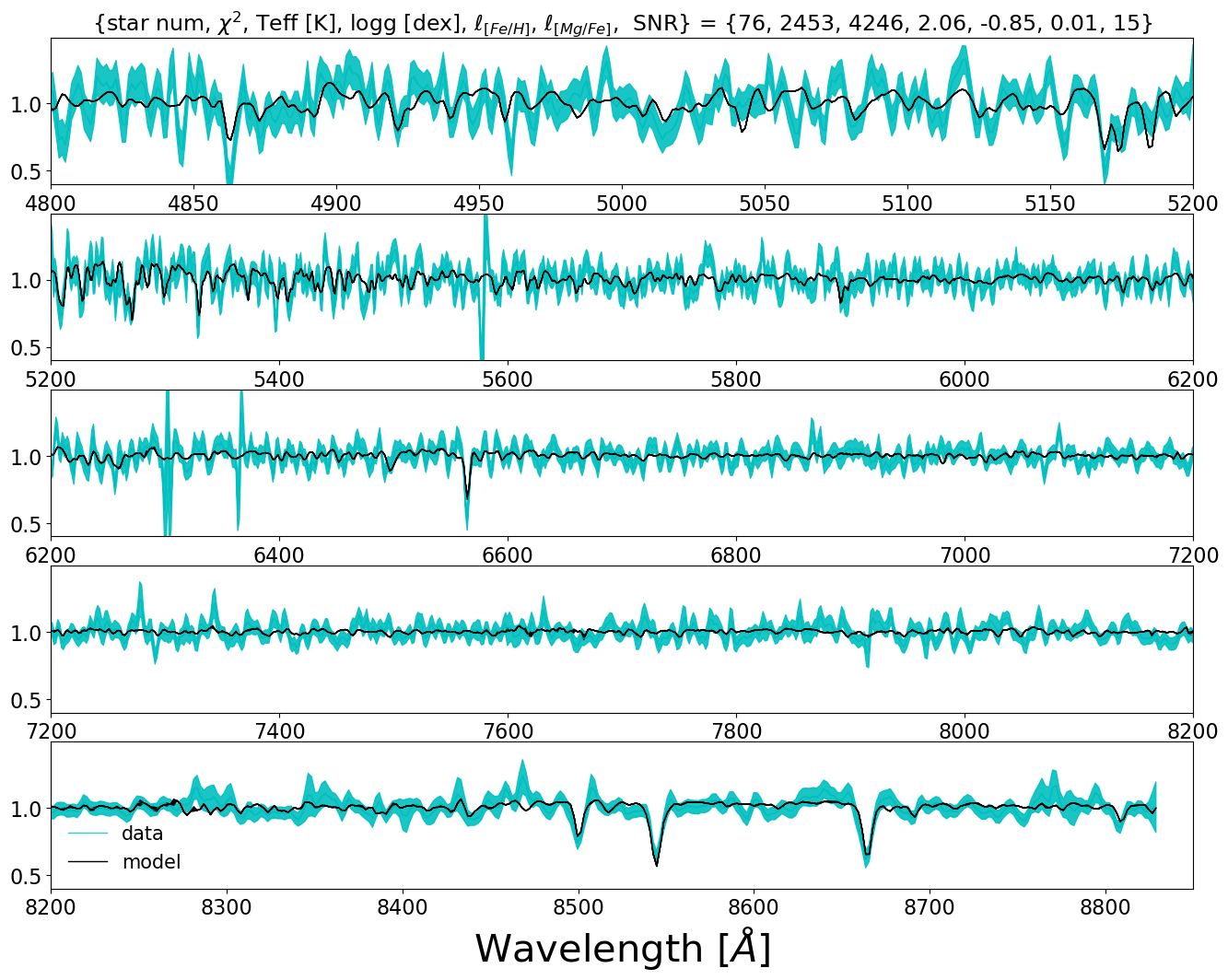}
\caption{Two example fits of the model (black) and data (cyan) for stars with 15 $<$ SNR  $<$ 20. Summary labels are indicated at top.}
\label{fig:specthree}
\end{figure*}

\begin{figure*}
\centering
\includegraphics[scale=0.36]{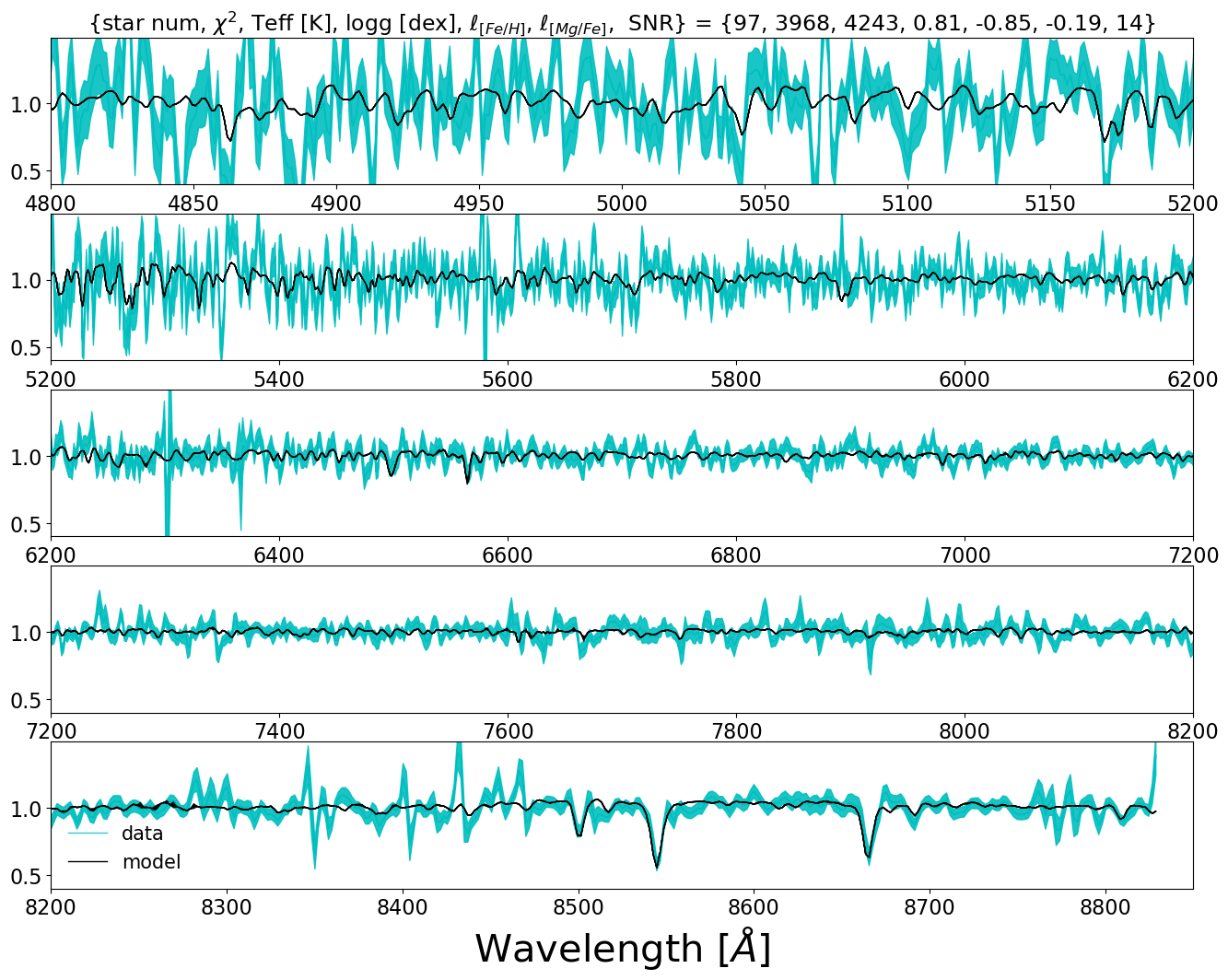}
\includegraphics[scale=0.36]{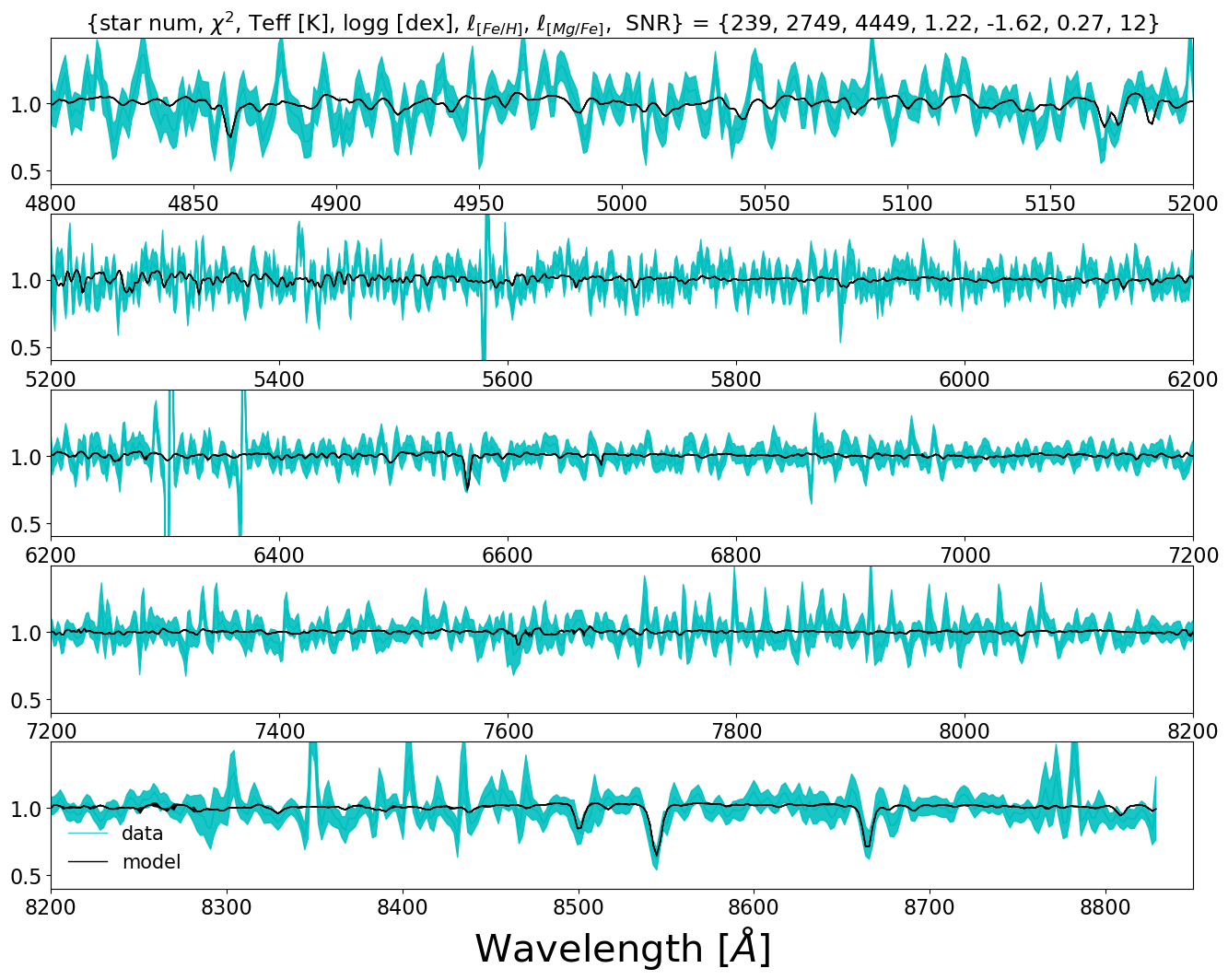}
\caption{Two example spectra and model fits for stars with 10 $<$ SNR $<$ 15; spectra in cyan and model in black. Note there are data reduction artefacts such as poorly subtracted tellurics, sky and cosmic rays.}
\label{fig:specfour}
\end{figure*}

Figures \ref{fig:specone}, \ref{fig:spectwo},  \ref{fig:specthree} and \ref{fig:specfour} show examples of spectra from 10 $<$ SNR $<$ 50. The model is in black and the data is in cyan. A subset of the parameters are included for each of the examples and the full set of labels for each is in the online table. 

%makeplot_loop.py

%makeplot_loop.py

\label{lastpage}
\end{document}